\documentclass[twocolumn]{aastex7}

\newcommand{\lya}{\textrm{Ly}\ensuremath{\alpha}}
\newcommand{\oiii}{[O\,{\sc III}]}

\usepackage{comment}
\usepackage{amsmath}
\usepackage{xcolor}
\definecolor{myblue}{HTML}{7130FC}

\begin{document}

\title{Probing Direct Contributions of Galaxies and AGN to Cosmic Reionization in a Quasar Field J0226$+$0302 with JWST NIRCam and NIRSpec} 

\suppressAffiliations

\correspondingauthor{Xiangyu Jin}
\email{jxiangyu@umich.edu}

\author[orcid=0000-0002-5768-738X,sname='Jin']{Xiangyu Jin}
\affiliation{Department of Astronomy, University of Michigan, 1085 S. University Ave., Ann Arbor, MI 48109, USA}
\affiliation{Steward Observatory, University of Arizona, 933 N Cherry Ave, Tucson, AZ 85721, USA}
\email{jxiangyu@umich.edu}  

\author[0000-0001-5287-4242]{Jinyi Yang}
\affiliation{Department of Astronomy, University of Michigan, 1085 S. University Ave., Ann Arbor, MI 48109, USA}
\email{}

\author[0000-0002-7633-431X]{Feige Wang}
\affiliation{Department of Astronomy, University of Michigan, 1085 S. University Ave., Ann Arbor, MI 48109, USA}
\email{}

\author[0000-0001-6874-1321]{Koki Kakiichi}
\affiliation{Cosmic Dawn Center (DAWN), Niels Bohr Institute, University of Copenhagen, Jagtvej 155, Copenhagen N, DK-2200, Denmark}
\email{}

\author[0000-0003-3310-0131] {Xiaohui Fan}
\affiliation{Steward Observatory, University of Arizona, 933 N Cherry Ave, Tucson, AZ 85721, USA}
\email{}

\author[0000-0002-6021-7020]{Enrico Garaldi}
\affiliation{Kavli IPMU (WPI), UTIAS, The University of Tokyo, Kashiwa, Chiba 277-8583, Japan}
\affiliation{Center for Data-Driven Discovery, Kavli IPMU (WPI), UTIAS, The University of Tokyo, Kashiwa, Chiba 277-8583, Japan}
\email{}

\author[0000-0002-6184-9097]{Jaclyn B. Champagne}
\affiliation{Steward Observatory, University of Arizona, 933 N Cherry Ave, Tucson, AZ 85721, USA}
\email{}

\author[0000-0003-2344-263X]{George D. Becker}
\affiliation{Department of Physics \& Astronomy, University of California, Riverside, CA 92521, USA}
\email{}

\author[0000-0003-3307-7525]{Yongda Zhu}
\affiliation{Steward Observatory, University of Arizona, 933 N Cherry Ave, Tucson, AZ 85721, USA}
\email{}

\author[0000-0003-0111-8249]{Yunjing Wu}
\affiliation{Kavli Institute for the Physics and Mathematics of the Universe (WPI), The University of Tokyo Institutes for Advanced Study, The University of Tokyo, Kashiwa, Chiba 277-8583, Japan}
\email{yunjing.wu@ipmu.jp}

\author[0000-0001-9191-9837] {Marianne Vestergaard}
\affiliation{DARK, Niels Bohr Institute, University of Copenhagen, Jagtvej 155, Copenhagen N, DK-2200, Denmark}
\affiliation{Steward Observatory, University of Arizona, 933 N Cherry Ave, Tucson, AZ 85721, USA}
\email{}

\author[orcid=0000-0002-3211-9642]{Huanqing Chen}
\affiliation{Department of Science, Augustana Campus,
University of Alberta,
Camrose, AB T4V2R3, Canada}
\email{huanqing.chen@ualberta.ca}

\author[0000-0003-3693-3091]{Valentina D’Odorico}
\affiliation{INAF - Osservatorio Astronomico, via G.B. Tiepolo, 11, I-34143 Trieste, Italy}
\affiliation{IFPU - Institute for Fundamental Physics of the Universe, via Beirut 2, I-34151 Trieste, Italy}
\email{}

\author[0000-0003-2895-6218]{Anna-Christina Eilers}
\affiliation{Department of Physics, Massachusetts Institute of Technology, Cambridge, MA 02139, USA}
\affiliation{MIT Kavli Institute for Astrophysics and Space Research, Massachusetts Institute of Technology, Cambridge, MA 02139, USA}
\email{}

\author[0000-0002-5721-0709]{Jiamu Huang}
\affiliation{Department of Physics, University of California, Santa Barbara, CA 93106, USA}
\email{}

\author[0000-0003-1470-5901]{Hyunsung D. Jun}
\affiliation{Department of Physics, Northwestern College, 101 7th St. SW, Orange City, IA 51041, USA}
\affiliation{School of Physics, Korea Institute for Advanced Study, 85 Hoegiro, Dongdaemun-gu, Seoul 02455, Republic of Korea}
\email{}

\author[0000-0001-6251-649X]{Mingyu Li}
\affiliation{Department of Astronomy, Tsinghua University, Beijing 100084, China}
\email{lmy22@mails.tsinghua.edu.cn}
\email{}

\author[0000-0003-4924-5941]{Maria Pudoka}
\affiliation{Steward Observatory, University of Arizona, 933 N Cherry Ave, Tucson, AZ 85721, USA}
\email{}

\author[orcid=0000-0003-0747-1780]{Wei Leong Tee}
\affiliation{Department of Astronomy and Astrophysics, The Pennsylvania State University, 525 Davey Laboratory, University Park, PA 16802, USA}
\email{wmt5159@psu.edu}

\author[0000-0002-5367-8021]{Minghao Yue}
\affiliation{Steward Observatory, University of Arizona, 933 N Cherry Ave, Tucson, AZ 85721, USA}
\email{}

\author[0000-0002-0123-9246]{Huanian Zhang} 
\affiliation{Department of Astronomy, Huazhong University of Science and Technology, Wuhan, Hubei 430074, China}
\email{}

\author[0000-0002-3983-6484]{Siwei Zou}
\affiliation{Chinese Academy of Sciences South America Center for Astronomy, National Astronomical Observatories, CAS, Beijing 100101, China}
\affiliation{Departamento de Astronom\'ia, Universidad de Chile, Casilla 36-D, Santiago, Chile}
\email{zousw@nao.cas.cn}

\begin{abstract}

We present JWST Cycle 2 NIRCam and NIRSpec observations in a quasar field J0226+0302 at $z=6.5412$ to probe the direct connections between the intergalactic medium (IGM), galaxies, and AGN during reionization. This field was previously observed by the JWST ASPIRE program, detecting eight [OIII]-emitting galaxies at $5.3<z<6.4$ with a single NIRCam pointing. 
Using new NIRCam and NIRSpec observations, we identify 65 additional line-emitting galaxies at $5.3<z<6.4$. The IGM-galaxy cross-correlation function shows a $\sim2\sigma$ excess IGM transmission at $\sim10$--$40$~cMpc from galaxies, suggesting a significant contribution from regions traced by star-forming galaxies to the local ionizing background. The IGM–galaxy cross-correlation function in this field is consistent with THESAN simulations for an IGM neutral fraction of $5\%$--$7\%$ and an average galaxy ionizing photon escape fraction $f_{\rm esc}$ of $6\%$, although a single field is subject to substantial cosmic variance and cannot constrain these quantities independently.
Among 49 line-emitting galaxies observed by NIRSpec, we identify four AGN with broad H$\alpha$ emission lines, resulting in an AGN fraction of $(8\pm4)$\%. From the IGM effective optical depth and the IGM-AGN cross-correlation function, we find that the IGM transmission is higher within 5~$h^{-1}\;\!{\rm cMpc}$ of four AGN than around the majority of \oiii\ emitters. We interpret the excess transmission as resulting from the local radiation enhancement by the AGN, and estimate $f_{\rm esc}\sim50\%$--$100\%$, based on our illustrative model for the IGM-AGN cross-correlation function. Future JWST NIRSpec observations in quasar fields will yield a more constraining IGM-AGN cross-correlation function, providing further insights into the roles of galaxies and AGN in reionization. 

\end{abstract}

\keywords{\uat{Reionization}{1383} --- \uat{Quasars}{1319} --- \uat{Intergalactic medium}{813} --- \uat{Quasar absorption line spectroscopy}{1317} --- \uat{Lyman alpha forest}{980} ---\uat{High-redshift galaxies}{734} --- \uat{Active galactic nuclei}{16}}


\section{Introduction} \label{sec:intro}

Cosmic reionization was the last major phase transition of neutral hydrogen in the intergalactic medium (IGM), driven by the first luminous sources. Cosmic reionization therefore encodes the collective radiative history of the first luminous sources. Studies of cosmic reionization strongly inform the early structure formation, galaxy formation, and how early luminous sources shaped the IGM and influenced galaxy evolution. 
With extensive observational studies over two decades 
\cite[for reviews, see][]{Fan2006ARAA,Fan2023ARAA,Becker2015PASA,Stark2016ARAA,Ouchi2020ARAA,Robertson2022ARAA}, the timeline of the second half of reionization has been moderately constrained, with a midpoint of reionization at $z\sim7-8$ \cite[e.g.,][]{Mason2018ApJ,Planck2020AA}, a late end at $z<6$ \cite[e.g.,][]{Yang2020ApJ,Zhu2021ApJ,Zhu2024MNRAS,Jin2023ApJ}, and a tail down to $z\sim5.3$ \cite[e.g.,][]{Zhu2021ApJ,Bosman2022MNRAS}. Although the onset of reionization is loosely constrained \cite[e.g.,][]{Cain2025ApJ}, \citet{Witstok2025Natur} report a detection of a Lyman $\alpha$ (Ly$\alpha$) emission line at $z=13$, implying that reionization started just 0.3~Gyr after the Big Bang. 

However, the dominant sources of cosmic reionization still remain poorly constrained. Prior to JWST, star-forming galaxies were believed to be the dominant sources of reionization \cite[e.g.,][]{Finkelstein2019ApJ,Naidu2020ApJ}, with luminous quasars only contributing a minor fraction to reionization \cite[e.g.,][]{Jiang2022NatAs}. On the other hand, JWST has uncovered an abundant and mysterious population of ``little red dots" at $z>5$, often characterized by a ``V-shaped" spectral energy distribution, compact morphology, and broad emission lines \cite[e.g.,][]{Harikane2023ApJ,Kocevski2023ApJ,Fujimoto2024ApJ,Kokorev2024ApJ,Labbe2025ApJ,Lin2024ApJ,Lin2026ApJ,Matthee2024ApJ,Maiolino2024AA,Zhang2025arXiv}, potentially tracing active galactic nuclei (AGN) activity. 
It remains debated whether those JWST-discovered AGN can contribute significantly to reionization (\citealt{Harikane2023ApJ,Fujimoto2024ApJ,Grazian2024ApJ,Madau2024ApJ}, but see also \citealt{Atek2024Natur,Matthee2024ApJ,Jiang2025arXiv,Kakiichi2025arXiv,Asthana2025MNRAS}). More observations are thus required to investigate the contribution of galaxies and AGN to reionization. 

Direct measurements of galaxy Lyman Continuum flux during reionization are nearly impossible due to the absorption of neutral hydrogen in the foreground IGM. The quasar Ly$\alpha$ forest provides an efficient, alternative way of mapping the foreground IGM transmission and probing direct connections with ionizing sources whose ionized regions intersect the quasar sightline (i.e., transverse proximity effect). \citet{Gallerani2008MNRAS} detect the direct flux enhancement in the quasar Ly$\alpha$ forest of the quasar SDSS\;J1148+5251 at $z=6.42$ due to the existence of a foreground quasar RD J1148+5253 at $z\sim5.7$ \cite[$m_{\rm 1450\;\!{\rm \AA}}=23.1$, ][]{Leipski2014ApJ}. The transverse proximity effect of the quasar SDSS J0100$+$2802 is also detected in the Ly$\alpha$ forest of background galaxies \citep{Eilers2025arXiv}. 

\citet{Kakiichi2018MNRAS} proposed using the cross-correlation between the quasar Ly$\alpha$ forest and the spatial location of galaxies (hereafter IGM-galaxy cross-correlation function) in the quasar fields to constrain sources of reionization,  because an excess of IGM transmission is expected around galaxies. 
Using Lyman-break galaxies (LBGs) and Lyman $\alpha$ emitters (LAEs) identified in the quasar fields from ground-based observations, \citet{Kakiichi2018MNRAS} and \citet{Meyer2019MNRAS,Meyer2020MNRAS} find excess transmission at distances greater than $\sim10$~comoving megaparsec (cMpc) around galaxies, but the scale and the significance of the excess transmission depend on the selection of galaxy tracers. In addition, the IGM-galaxy cross-correlation function also suffers from large uncertainties using a limited number of both quasar fields and galaxies \citep{Meyer2020MNRAS}. 

JWST has opened a new era for the identification of high-redshift galaxies in quasar fields, allowing the study of the direct connection between quasar Ly$\alpha$ forest and galaxies. In JWST Cycle 1, the program ``Emission-line galaxies and Intergalactic Gas in the Epoch of Reionization"(EIGER, PID: 1243, PI: S. Lilly) performed galaxy redshift surveys in six quasar fields at $z>6$ with NIRCam Wide Field Slitless Spectroscopy (WFSS) in the F356W filter in a $2\times2$ mosaic, and the program ``A SPectroscopic Survey of Biased Halos in the Reionization Era" (ASPIRE, PID: 2078, PI: F. Wang) observed 25 quasar fields at $6.5<z<6.8$ with a single NIRCam/WFSS pointing also in the F356W filter. Both programs enable the analysis of IGM-galaxy cross-correlation function. 
With the full EIGER sample of six quasar fields, \citet{Kashino2025arXiv} report enhanced IGM transmission at distances of 5--20~cMpc from \oiii\ emitters at $z>5.7$ relative to the average IGM transmission, whereas no significant enhancement is detected at $z<5.7$. In contrast, \citet{Kakiichi2025arXiv} detect enhanced IGM transmission at distances of 20--40~cMpc from \oiii\ emitters identified in five ASPIRE quasar fields. Using 11 ASPIRE quasar fields, \citet{Jin2024ApJ} also report enhanced IGM transmission around \oiii\ emitters on scales of $\gtrsim25h^{-1}\;\!{\rm cMpc}$ by measuring the IGM effective optical depth centered on the \oiii\ emitters. However, most of these studies in quasar fields only observed galaxies with NIRCam WFSS in F356W. For $z\sim6$ galaxies, the NIRCam F356W filter covers only the H$\beta$ and \oiii\ emission lines. 
This narrow wavelength coverage prevents us from confirming the nature of \oiii\ emitters as ionizing sources and further characterizing their properties.  

In this work, we present NIRCam/WFSS and NIRSpec/microshutter array (MSA) observations from a JWST Cycle 2 GO program 3325 (PIs: F. Wang and J. Yang), which observed two quasar fields J0226$+$0302 and J0305$-$3150 with NIRCam/WFSS, NIRCam/imaging, NIRSpec/MSA, and NIRSpec/IFU. NIRCam/WFSS, NIRCam/imaging and NIRSpec/MSA enable the studies of large-scale structures in the quasar fields. NIRCam/WFSS and imaging provide an efficient source identification through \oiii\ emission lines and NIRSpec/MSA provides a wider wavelength coverage and a higher sensitivity for characterizing the source properties. The central quasars were also observed with NIRSpec/IFU to investigate the small-scale environment around the quasar. Both quasar fields were included as part of the JWST Cycle 1 program ASPIRE \cite[GO 2078;][]{Wang2023ApJ, Yang2023ApJ}, and both quasars have significant overdensity of [O III]-emitting galaxies $\delta_{\rm gal, [O\;III]}\gtrsim15$ at the quasar redshifts \citep{Wang2023ApJ,Wang2026arXiv,Champagne2025ApJa,Champagne2025ApJb}. In this work, we focus on galaxies in the quasar field J0226$+$0302 \cite[hereafter J0226, $z_{\rm QSO}=6.5412$, ][]{Banados2016ApJS}, whose redshift is at $5.4<z<6.4$, in the quasar foreground and within the quasar \lya\ forest.
All galaxies included in our analysis have secure redshift determination through the detection of rest-frame optical emission lines, either from NIRCam/WFSS or NIRSpec/MSA. 

This paper is organized as follows: In Section \ref{sec:2}, we present the data reduction procedures for JWST NIRCam and NIRSpec observations in the J0226 field, and ground-based spectroscopy covering the quasar Ly$\alpha$ forest. In Section \ref{sec:3}, we present our analysis of identifying sources from NIRCam and NIRSpec, measurements of their properties through photometry and spectroscopy, and identification of AGN. The IGM-galaxy cross-correlation function is presented in Section \ref{sec:results}. We discuss the determination of ionizing photon production efficiency in \ref{sec:ion} and the AGN transverse proximity effect in Section \ref{sec:discussion}. We summarize the paper in Section \ref{sec:summary}. Throughout this paper, we adopt a flat $\Lambda$CDM cosmology with $H_{\rm 0}=70~{\rm km\;\!s^{-1}\;\!Mpc^{-1}}$, $\Omega_{\rm m}=0.3$ and $\Omega_{\Lambda}=0.7$. All quoted magnitudes are in the AB system \citep{Oke1983ApJ}.

\section{Data Reduction} \label{sec:2} 
\subsection{JWST NIRCam Imaging and WFSS} \label{sec:nircam}

J0226 was observed on August 10, 2022 as part of the ASPIRE program (GO 2078) with a single NIRCam WFSS pointing in F356W with a simultaneous F200W imaging, and direct imaging in F115W and F356W. The WFSS in F356W provides a wavelength coverage of $3.14-3.98~\mu$m, covering \oiii$\lambda\lambda$4960,5008 emission lines at $z\sim5.3-7.0$, with a spectral resolution of $R\sim1600$. The exposure time of the WFSS observation is 2834.5~s, and the exposure time of imaging is 1417.3~s, 
including the direct imaging and out of field imaging. In Cycle 2 (GO 3325), J0226 was observed during Aug 29 $-$ Sep 1 2023 with five more NIRCam pointings. In total, six pointings form a $2\times3$ mosaic centered at the quasar, covering a field of view (FoV) of $4.3' \times7.3'$. Each pointing has the same exposure series in imaging and WFSS, and the same visiting position angle as the NIRCam observations of J0226 in Cycle 1. We refer readers to \citet{Wang2023ApJ} and \citet{Yang2023ApJ} for details regarding the NIRCam WFSS and imaging data reduction. We follow the procedure in \citet{Wang2026arXiv} to identify \oiii\ emitters. Selected \oiii\ emitters were visually inspected by our team members. We identify 121 bright \oiii\ emitters with at least two emission lines detected in both 1D and 2D spectra, as well fainter \oiii\ emitters which only show line detections in either 1D or 2D spectra. Bright \oiii\ emitters and faint \oiii\ emitter candidates are primary targets for NIRSpec/MSA observations.

Figure \ref{fig:pointing} shows all existing NIRCam pointings in the J0226 field. The single NIRCam pointing from Cycle 1 ASPIRE program is plotted in purple, and five NIRCam pointings obtained from GO 3325 program are plotted in green. 

\subsection{JWST NIRSpec/MSA} 
\label{sec:nirspec}

\begin{figure}
    \centering
     \includegraphics[width=1.0\linewidth]{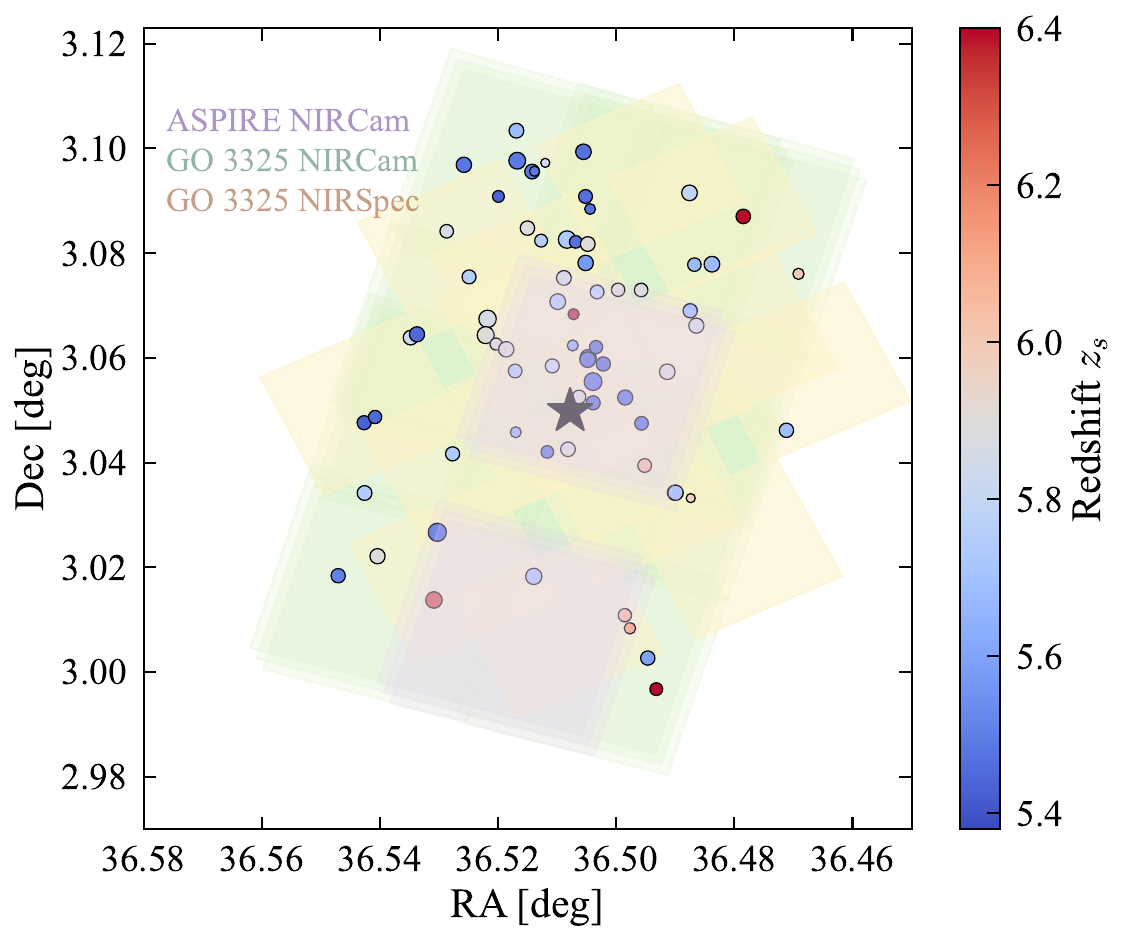}
    \caption{The pointings of existing JWST NIRCam (purple -- ASPIRE GO 2078; green -- GO 3325) and NIRSpec/MSA observations (yellow) in the J0226 field. The quasar at $z=6.5412$ is marked with a black star. All spectroscopically confirmed galaxies at $5.38<z<6.4$ are denoted by dots. The size of the markers is scaled with $M_{\rm 1500}$ of the source, and the marker is color-coded by the source redshift $z_s$.}
    \label{fig:pointing}
\end{figure}

Our targets for NIRSpec/MSA observations are WFSS-selected \oiii\ emitters and LBGs with $z_{phot}>5.3$. LBG candidates are selected to have SNR$<$2 in filters shortward of F115W (which included archival ground-based Sloan $g$, $i$ and $z$) and then were visually inspected to eliminate compact stellar candidates and image artifacts. We fit each candidate with EAZY and further require 70\% of the EAZY $P(z)$ to lie between $5.3<z_{phot}<7$. LBG candidates are prioritized for MSA slits in brightness order. 
We use the G395M grating and F290LP blocking filter, resulting in a wavelength range of $2.87-5.10~\mu$m with a spectral resolution $R\sim1000$. In the J0226 field, we have four MSA pointings in total. For each pointing, we use a NIRSIRS2 readout and employ a 3-shutter slitlet with a 3-point nodding pattern. The total exposure time for each pointing is 3545.1s. Three pointings were observed on Dec 26, 2023 and one pointing was observed on Jan 25, 2024. 
The MSA pointings are shown in yellow in Figure \ref{fig:pointing}. With four NIRSpec/MSA pointings, we observe 512 sources in total, including 82 bright WFSS-selected \oiii\ emitters with at least two lines detected in both 1D and 2D. Filler targets include WFSS-selected faint \oiii\ emitter candidates and line emitters at other redshifts, and photometrically selected LBGs.  

For NIRSpec/MSA data reduction, we use the JWST Calibration pipeline \texttt{v1.16.0} and Calibration References Data System (CRDS) mapping \texttt{jwst\_1298.pmap}. We use \texttt{Detector1Pipeline} to perform the detector-level corrections, including flat field, dark current, and bias to generate rate files. 
During the Stage 1, we employ \texttt{clean\_flicker\_noise} to remove the $1/f$ noise. For the jump step, we follow CEERS NIRSpec reduction parameters\footnote{\url{https://web.corral.tacc.utexas.edu/ceersdata/DR07/NIRSpec/README_NIRSpec_DR0.7.txt}} to improve the rejection of cosmic rays and snowballs. 
We use \texttt{msaexp} \texttt{v.0.9.7} \citep{Brammer2022zndo...7299500B} to extract 2d spectra for each slit and each exposure, correcting slit losses, path losses, and bar shadows. We then use \texttt{msaexp} to find the object trace in each exposure, combine the rectified 2d spectra, and extract the 1d spectra.

\subsection{Ground-based Optical Spectroscopy}

J0226 is included in the E-XQR-30 program \citep{Dodorico2023MNRAS}, and was observed with X-Shooter on the Very Large Telescope (VLT) with a total exposure time of 6.5 hours in the visual (VIS) arm and a spectral resolution of $R\sim10700$. We downloaded the VIS spectrum of J0226 from the E-XQR-30 Github repository\footnote{\url{https://github.com/XQR-30/Spectra}}. The VIS spectrum is rebinned to a pixel size of 10~${\rm km\;\!s^{-1}}$. In our prior work \citep{Jin2024ApJ,Kakiichi2025arXiv}, the same spectrum was used in the analysis. We briefly summarize the procedure to calculate the transmitted flux in the quasar Ly$\alpha$ forest. Following \citet{Jin2023ApJ}, we first mask the spectral pixels which are likely influenced by the sky emission lines and we fit the quasar spectrum at rest-frame $1245-1285~{\rm \AA}$ and $1310-1380~{\rm \AA}$ with a power-law with a spectral index $\alpha_{\lambda}$ of $-1.5$ (i.e., $f_{\lambda}\propto \lambda^{\alpha_{\lambda}}$), and adopt a broken power-law model to calculate the intrinsic quasar continuum flux: $\alpha_{\lambda}=-1.5$ at $\lambda>1000~{\rm \AA}$, and $\alpha_{\lambda}=-0.59$ at $\lambda\leq1000~{\rm \AA}$ \citep{Shull2012ApJ}. 
We calculate the IGM Ly$\alpha$ forest transmission as $T_{\rm Ly\alpha}=F_{\rm observed}/F_{\rm continuum}$, where $F_{\rm observed}$ is the observed flux in the quasar Ly$\alpha$ forest and $F_{\rm continuum}$ is the intrinsic quasar continuum flux.

\section{Analysis} \label{sec:3}

\subsection{Spectroscopic Redshift Determination}

To probe the connection between quasar Ly$\alpha$ forest and galaxies, our targets of interest are located in the foreground of the quasar ($z_{\rm QSO}=6.5412$), covered by the quasar Ly$\alpha$ forest. We choose sources at $z<6.4$ to avoid contaminants from the quasar proximity zone. H$\beta$ and \oiii$\lambda\lambda$4960,5008 emission lines at $z\sim5.3-7.0$ are covered by the NIRCam F356W WFSS. If such sources are selected as NIRSpec/MSA targets, their H$\beta$, \oiii$\lambda\lambda$4960,5008 and H$\alpha$ emission lines are expected to fall within the NIRSpec G395M/F290LP wavelength window. For WFSS sources, we choose those \oiii\ emitters which are detected in both 1D and 2D spectra. 

To determine the redshift of all NIRSpec/MSA sources, we adopt an emission line template of a star-forming galaxy, including [\ion{O}{2}]$\lambda$3726, H$\gamma$, H$\beta$, \oiii$\lambda\lambda$4960,5008 and H$\alpha$ emission lines. We cross-correlate the emission line template with all 1D spectra, and identify the redshift resulting from the maximum correlation as the cross-correlation redshift ($z_{\rm xcf}$). We also use \texttt{fit\_redshift} included in \texttt{msaexp} to identify the redshift of the source ($z_{\rm msaexp}$). We visually inspect all MSA sources with redshift determination from cross-correlation function ($z_{\rm xcf}$), msaexp ($z_{\rm msaexp}$), and NIRCam WFSS ($z_{\rm WFSS}$, if available) to determine the spectroscopic redshift ($z_{\rm s}$). 

For sources observed by NIRSpec/MSA, we choose those sources in our analysis with at least two emission lines matching H$\beta$, \oiii$\lambda\lambda$4960,5008, H$\alpha$ at $z_{\rm xcf}$ or $z_{\rm msaexp}$. For sources with \oiii\ emission lines detected in both 2D and 1D WFSS spectra, because WFSS achieves a higher spectral resolution ($R\sim1600$) than NIRSpec ($R\sim1000$), we adopt the redshift measured from WFSS. Among 82 WFSS-selected \oiii\ emitters observed by NIRSpec/MSA, we confirm 80 \oiii\ emitters with at least two emission lines at the grism redshift $z_{\rm WFSS}$. The other two WFSS-selected \oiii\ emitters do not show emission lines in the NIRSpec spectra. We examine the NIRCam WFSS observations and find the \oiii\ emission of these two \oiii\ emitters duplicate emission lines of other identified \oiii\ emitters. As a result, we remove these two \oiii\ emitters from our analysis. For sources with \oiii\ emission lines detected in only 1D or 2D WFSS spectra, if there are two emission lines detected in the NIRSpec/MSA spectrum matching H$\beta$, \oiii$\lambda\lambda$4960,5008, H$\alpha$ at the corresponding $z_{\rm WFSS}$, we determine the source redshift using $z_{\rm WFSS}$. We confirm 20 faint WFSS-selected \oiii\ emitter candidates with emission lines at $z_{\rm WFSS}$ in the NIRSpec/MSA observations. For photometrically-selected LBGs, the source redshift is determined based on the NIRSpec/MSA spectrum through $z_{\rm xcf}$ or $z_{\rm msaexp}$. We identify seven line-emitting galaxies at $z>5.3$ among the LBG sample. Two galaxies are at $z=5.31$, where the NIRCam F356W WFSS sensitivity decreases, potentially leading to missed line detections in WFSS. The other five galaxies have an \oiii$\lambda$5008 flux less than $2\times10^{-18}\;\!{\rm erg\;\!s^{-1}\;\!cm^{-2}}$ in the NIRSpec spectrum, missed by the shallow depth of the WFSS observations. 

In total, from NIRCam and NIRSpec spectroscopy, we identify 146 sources at $z>5.3$.  
73 of these sources are located in the foreground of the quasar, at $z<6.4$. 
Table \ref{tab:spec-z} summarizes the information of 73 spectroscopically confirmed sources at $z<6.4$, including coordinates, the spectroscopic redshift, and their NIRCam photometry in F115W. Figure \ref{fig:pointing} shows the 2D spatial distribution of spectroscopically confirmed sources. 

\begin{deluxetable*}{ccccccc}\centering
\tabletypesize{\scriptsize}
\tablecaption{Summary of Spectroscopically Confirmed Sources in J0226 Quasar Field}
\tablewidth{\textwidth}
\tablehead{\colhead{Source ID} &
\colhead{RA} & \colhead{Dec} & \colhead{$z_{s}$} & \colhead{$m_{\rm F115W}$}& \colhead{$z_s$ Source} &\colhead{Broad H$\alpha$ Flag}}
\colnumbers
\startdata
90  &  36.5427  &  3.0476  &  5.430  &  27.64  &  wfss & 0  \\ 
345  &  36.5408  &  3.0487  &  5.448  &  27.82  &  wfss+msa & 0  \\ 
438  &  36.5258  &  3.0969  &  5.480  &  27.02  &  wfss & 0  \\ 
594  &  36.5287  &  3.0842  &  5.855  &  27.67  &  msa & 0  \\ 
632  &  36.5348  &  3.0639  &  5.834  &  27.24  &  wfss+msa & 1  \\ 
793  &  36.5337  &  3.0645  &  5.448  &  26.96  &  wfss & 0  \\ 
836  &  36.5426  &  3.0342  &  5.746  &  27.34  &  wfss & 0  \\ 
914  &  36.5471  &  3.0184  &  5.506  &  27.41  &  wfss & 0  \\ 
1722  &  36.5169  &  3.1034  &  5.679  &  27.24  &  wfss & 0  \\ 
1808  &  36.5249  &  3.0755  &  5.756  &  27.55  &  wfss+msa & 0  \\ 
$\cdots$ & $\cdots$ & $\cdots$ & $\cdots$ & $\cdots$ &  $\cdots$ \\
\enddata
\tablecomments{(1) Source ID; (2)-(3) Coordinates of identified sources; (4) The spectroscopic redshift $z_{s}$; (5) Forced photometry magnitude in F115W; (6) The instrument and observing mode used to derive each spectroscopic redshift, distinguishing NIRCam WFSS redshifts from NIRSpec micro-shutter array (MSA) redshifts; (7) Notes for identified broad H$\alpha$ emitter (FWHM $>1000~{\rm km\;\!s^{-1}}$), where ``$1$" indicates an identified broad H$\alpha$ emitter and ``$0$" otherwise. The full catalog will be available in the online version upon acceptance.}
\end{deluxetable*}\label{tab:spec-z}

\subsection{NIRSpec/MSA Spectrum Fitting}

For all NIRSpec sources with $z_{\rm s}$ within the quasar Ly$\alpha$ forest, we fit their NIRSpec spectrum to measure the emission line properties. We first mask pixels with negative fluxes (S/N $<-3$) in the NIRSpec spectrum. We use \texttt{lmfit} in Python to fit the rest-frame spectrum \citep{Newville_lmfit2016}. 
We fit the spectral regions around H$\beta$ and H$\alpha$ emission lines separately. For spectrum near $H\beta$, we use single Gaussian components for H$\beta$ and \oiii$\lambda\lambda$ 4960, 5008, and a flat continuum. We assume the same redshift and width for H$\beta$ and \oiii$\lambda\lambda$ 4960, 5008, and fix the flux ratio \oiii$\lambda5008/\lambda4960=2.98$ during the fitting \citep{Storey2000MNRAS}. 

In order to select AGN candidates with broad H$\alpha$ emission lines, we model the H$\alpha$ profile with two different models. One narrow H$\alpha$ model only contains narrow emission lines including H$\alpha$ and [\ion{N}{2}]$\lambda\lambda6585,6549$, and the other broad H$\alpha$ model includes an additional broad component for H$\alpha$ emission. We then assess which model provides a better fit to the data. For the narrow H$\alpha$ model, we use single Gaussian components for narrow H$\alpha$ and [\ion{N}{2}]$\lambda\lambda6585,6549$, and a flat continuum. We assume the same velocity for the narrow H$\alpha$ and [\ion{N}{2}]$\lambda\lambda6585,6549$ components, and keep the flux ratio [\ion{N}{2}]$\lambda6585/\lambda6549=3.049$ fixed \citep{Dojvinovic2023AdSpR}. The full width at half maximum (FWHM) is set to be in the range 100$~{\rm km\;\!s^{-1}}<$ FWHM $<$1000$~{\rm km\;\!s^{-1}}$ for narrow lines. For the broad H$\alpha$ model, we adopt a broad H$\alpha$ Gaussian component (FWHM$>$1000$~{\rm km\;\!s^{-1}}$) in addition to narrow H$\alpha$, [\ion{N}{2}] components, and the flat continuum. 

\begin{figure*}[!h]
    \centering
    \includegraphics[width=0.95\linewidth]{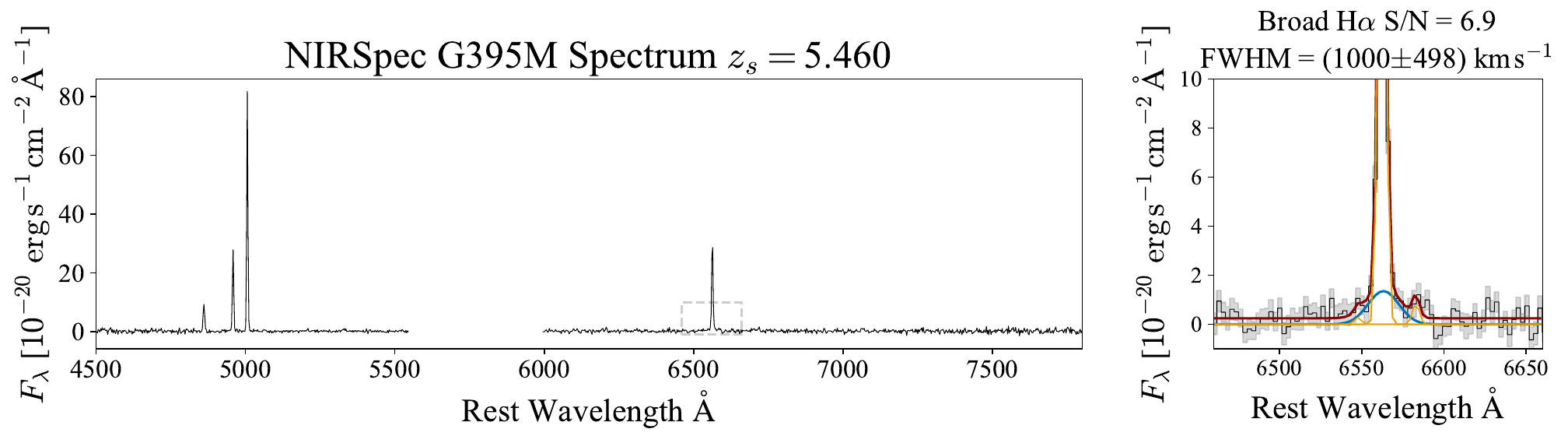}
    \includegraphics[width=0.95\linewidth]{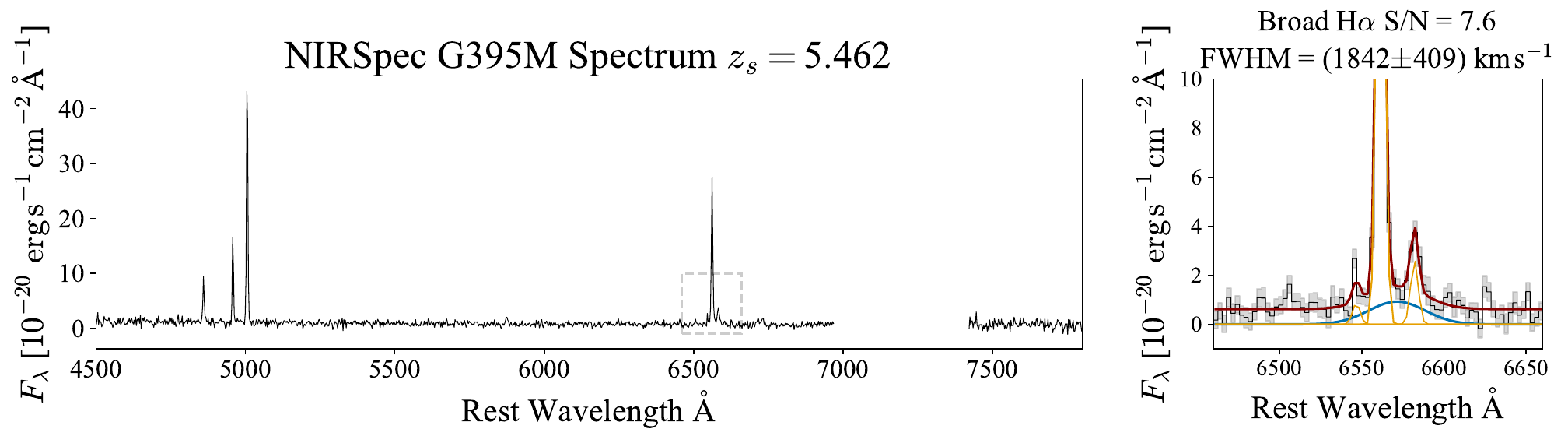}
    \includegraphics[width=0.95\linewidth]{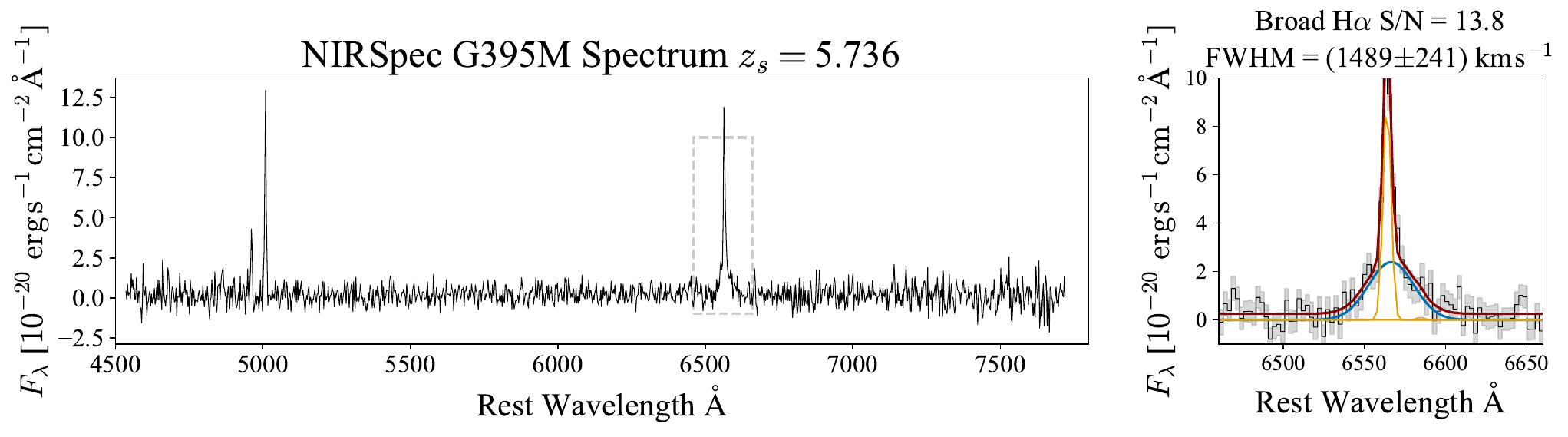}
    \includegraphics[width=0.95\linewidth]{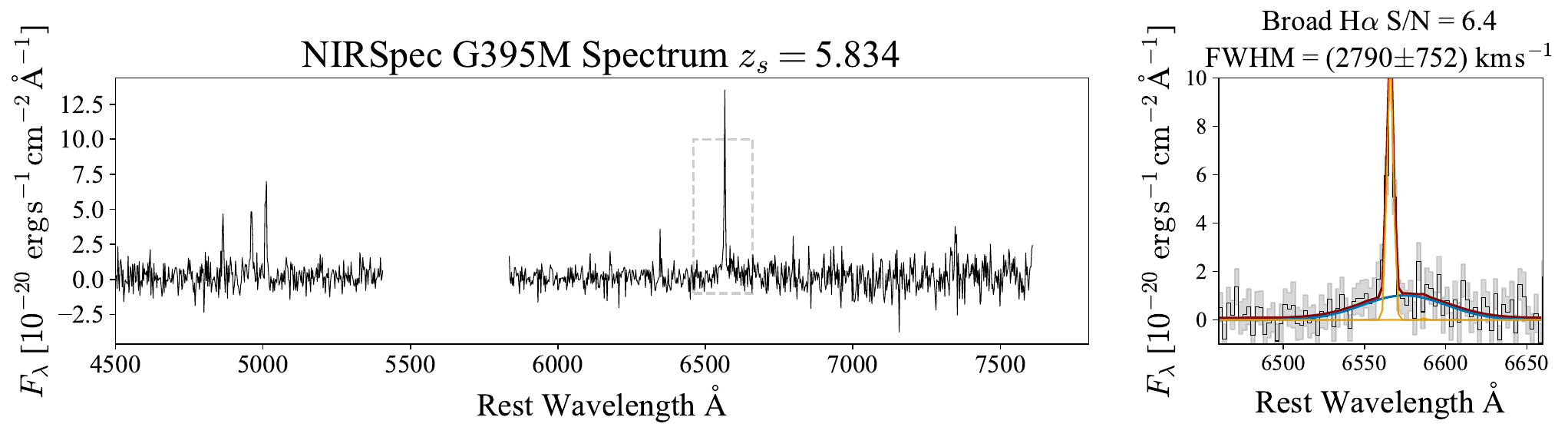}
    \caption{The NIRSpec spectra (left) and the H$\alpha$ profile (right) of the broad H$\alpha$ AGN candidates identified at $z<6.4$, ranked by the spectroscopic redshift $z_{\rm s}$. The best-fit model total is denoted by the red line. The best-fit broad H$\alpha$ line is plotted in blue. The best-fit narrow H$\alpha$ and [\ion{N}{2}] doublets are shown in orange.}
    \label{fig:AGN_candidate}
\end{figure*}

For each source, we compare the best-fit results of the narrow H$\alpha$ and the broad H$\alpha$ models, and identify broad H$\alpha$ emitters using the following criteria: (1) the broad H$\alpha$ model has a lower Bayesian Information Criterion (BIC) than the narrow H$\alpha$ model; (2) the broad H$\alpha$ model has a lower reduced chi-square $\chi_{\nu}^{2}$ than the narrow H$\alpha$ model; and (3) the integrated S/N of the best-fit broad H$\alpha$ component is greater than 5. Using these three criteria, we identify four broad H$\alpha$ emitters at $z<6.4$, among 49 line-emitting galaxies observed by NIRSpec/MSA. The corresponding broad-line AGN fraction is $(8\pm4)\%$, similar to the broad-line AGN fraction measured from NIRSpec data \citep{Harikane2023ApJ}, significantly higher than the broad-line AGN fraction of $1\%-2\%$ measured from NIRCam/WFSS data \citep{Matthee2024ApJ,Zhang2025arXiv,Lin2025arXiva}, highlighting the NIRspec's higher sensitivity in recovering the broad emission lines than NIRCam/WFSS. 
Figure \ref{fig:AGN_candidate} shows the NIRSpec spectrum of AGN candidates. The FWHM of the broad H$\alpha$ components derived from the best-fit model ranges from $(1000\pm498)~{\rm km\;\!s^{-1}}$ to $(2790\pm752)~{\rm km\;\!s^{-1}}$, and none of these four AGN candidates have broad H$\beta$ emission lines.

\subsection{Aperture Photometry} \label{sec:aper}
To estimate the depth of the observations, for each band, we mask all detected sources by \texttt{Source Extractor}, and use 5000 random apertures of 0.3\arcsec\, radius to measure the flux. Then we use the median absolute deviation of the flux within the random apertures to estimate the 1$\sigma$ uncertainty in the photometry. The 5$\sigma$ limiting magnitudes in F115W, F200W, and F356W are 27.4, 27.9, and 28.2.
Note that stacked NIRCam observations are deeper around the quasar than the outskirt regions (see Figure \ref{fig:pointing}). 
For all spectroscopically confirmed sources, we use \texttt{photutils} to measure their magnitude in F115W, F200W, and F356W. We use a circular aperture of 0.2\arcsec\, radius and perform the aperture flux correction using the encircled energy fraction of simulated JWST point spread function (PSF) \footnote{\url{https://jwst-docs.stsci.edu/files/216457060/216457084/1/1762453591836/Encircled_Energy_LW_ETCv2.txt}}. 
For every source, we generate 500 random apertures of 0.2\arcsec\ radius within a region of 1.8\arcsec centered on the source, and use the median absolute deviation of the flux within random apertures to estimate their local uncertainty. If the source has flux less than 2$\sigma$ uncertainty, we then use $2\sigma$ flux limit to estimate their magnitude. 
 
\begin{figure*}[!htb]
    \centering
    \includegraphics[width=0.95\linewidth]{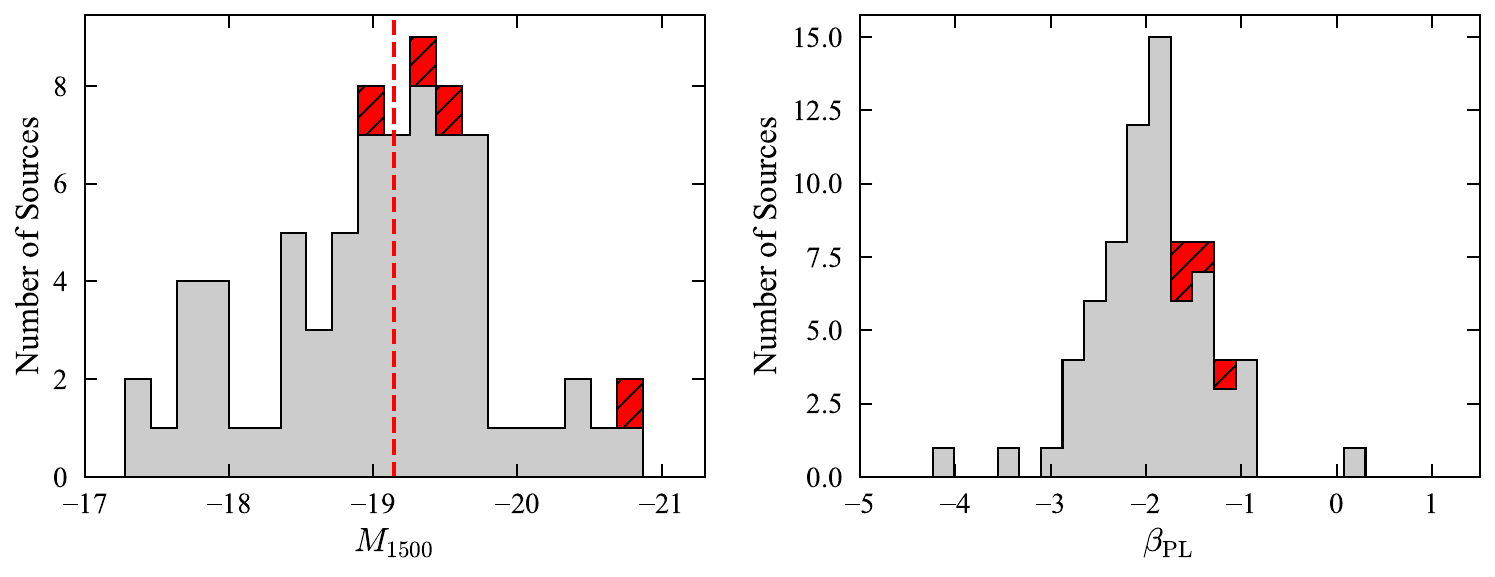}
    \caption{\textit{Left} -- The $M_{\rm 1500}$ distribution of all spectroscopically confirmed sources at $z<6.4$. The median $M_{\rm 1500}=-19.1$ is denoted by the vertical red dashed line. \textit{Right} -- The distribution of UV slopes $\beta_{\rm PL}$ derived through the photometric power-law fitting. In both panels, the distribution of narrow H$\alpha$ sources is plotted in a gray histogram, and the distribution of four broad H$\alpha$ emitters is plotted in a red hatched histogram.}
    \label{fig:Muv_dist}
\end{figure*}

Following \citet{Morales2025ApJ}, we calculate the UV continuum slope $\beta_{\rm PL}$ by fitting a power-law ($f_{\rm \lambda} \propto \lambda^{\beta}$) to F115W and F200W photometry. NIRCam F115W and F200W filters cover the rest-wavelength ranges of $1370-3460~{\rm \AA}$ for galaxies at $5.3<z<6.4$, well capturing their UV continuum. To estimate the UV magnitude $M_{\rm UV}$ of each source, we use the best-fit power-law to calculate their absolute magnitude at the rest-frame 1500~${\rm \AA}$ ($M_{\rm 1500}$). The left panel of Figure \ref{fig:Muv_dist} shows the $M_{\rm 1500}$ distribution of all sources with a median $M_{\rm 1500}$ of $-19.1$, and right panel shows the distribution of $\beta_{\rm PL}$. In both panel, the distributions of narrow H$\alpha$ sources are plotted in a gray histogram, and the distribution of broad H$\alpha$ emitters are plotted in a red hatched histogram. Four AGN display $\beta_{\rm PL}$ from $\sim-1.6$ to $\sim-1.3$, close to the quasar UV continuum slope $\beta\sim-1.5$ \citep{VandenBerk2001,Shull2012ApJ}. 

\section{Connections between Galaxies, CGM, and IGM} \label{sec:results}

\subsection{Spatial and Redshift Distribution of Spec-z Confirmed Sources and Their Direct Connections to IGM}

\begin{figure*}
    \centering
    \includegraphics[width=1.0\linewidth]{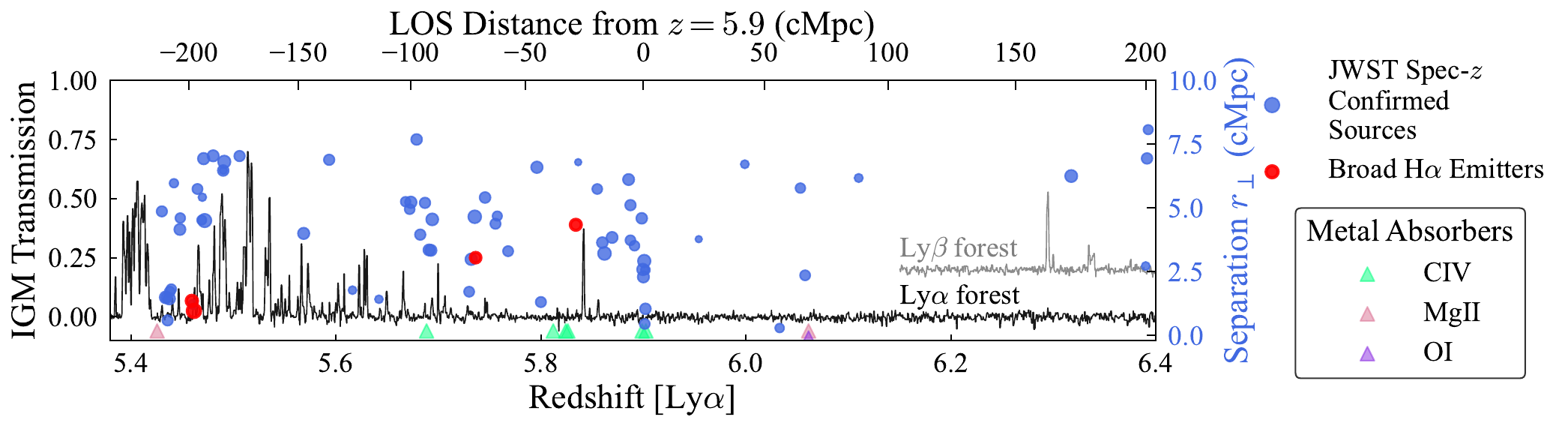}
    \caption{The quasar Ly$\alpha$ forest spectrum (black), Ly$\beta$ forest spectrum (gray), and spectroscopically confirmed sources (blue dots) from either JWST NIRCam/WFSS or NIRSpec/MSA. The Ly$\beta$ forest spectrum is vertically offset by 0.2 in IGM transmission for visualization purposes. The location of the source is denoted by their redshift and transverse distance to the quasar. The size of the symbol is scaled to $M_{\rm 1500}$, with a larger circle size representing a brighter $M_{\rm 1500}$. The sources showing a broad H$\alpha$ emission line are plotted with red dots. The redshifts of \ion{C}{4} (green), \ion{Mg}{2} (pink), and \ion{O}{1} (purple) absorbers identified by \citet{Davies2023MNRAS} are denoted by colored triangles on the x-axis.}
    \label{fig:los_plot}
\end{figure*}

To examine the relation between spectroscopically confirmed sources and the IGM, we show the Ly$\alpha$ forest spectrum and Ly$\beta$ forest spectrum of J0226, together with spectroscopically confirmed sources at $z<6.4$ in Figure \ref{fig:los_plot}. The spatial location of each source is denoted by its redshift $z_{\rm s}$ and transverse separation between the source and the quasar sightline in cMpc. The association between metal absorbers and galaxies is reported in Section \ref{subsec:metal}. 

This single quasar sightline displays distinct associations between galaxies and IGM transmission in the \lya\ forest at different redshifts. 
At $z=5.43-5.46$, the Ly$\alpha$ forest displays a strong absorption with a plethora of proximate galaxies detected close to the quasar sightline. Two AGN are identified at the gas absorption redshift of $z=5.46$, and transmission spikes are found at $\sim1-2$~cMpc of the AGN. At $z=5.5$, prominent Ly$\alpha$ transmission spikes are corresponding to an ``underdense" void of galaxies in the IGM, with no proximate galaxies detected within 5~cMpc of the quasar sightline. Such an association between the galaxy underdensity and the high IGM transmission has also been reported by \citep{Zhu2025arXiv} in the two most transparent sightlines at $z=5.7$. At $z=5.6-5.8$, many galaxies, including an AGN, are detected around a few transmission spikes, with faint galaxies detected close to transmission spikes in the transverse direction. 
An AGN is identified at $z=5.834$, and a strong Ly$\alpha$ transmission spike is detected at $z=5.84$, at $\sim4$~cMpc from the AGN. 

The Ly$\alpha$ forest of J0226 reaches nearly complete absorption at $z>5.9$ \cite[i.\,e., Gunn-Peterson trough,][]{GP1965ApJ}. At $z=6.3$, the Ly$\beta$ forest displays two transmission spikes with an \oiii\ emitting galaxy at $z=6.317$. Both Ly$\beta$ transmission spikes are located at $\sim10~$cMpc from the $z=6.317$ \oiii\ emitter. This Ly$\beta$ spike-\oiii\ emitter association has been previously reported in \citet{Kakiichi2025arXiv}, using a single NIRCam pointing from the ASPIRE program. However, with deeper and wider NIRCam observations in the J0226 field, no other galaxies are detected near the $z=6.317$ \oiii\ emitter.

\subsection{IGM-galaxy Cross-correlation Function}\label{sec:igm_galaxy_ccf}
 
To study the average IGM transmission as a function of distance from galaxies, we measure the IGM-galaxy cross-correlation function. In our calculation, we only use the quasar Ly$\alpha$ forest at rest-frame wavelength $>1040$\AA\, to remove the contaminants from Ly$\beta$ and \ion{O}{6} emissions. We also exclude the absorption redshift at $z>z_{\rm QSO}-0.1$ to remove the Ly$\alpha$ forest within the quasar proximity zone. This results in a Ly$\alpha$ absorption redshift range of $z=5.4-6.4$ used in our calculation. We calculate the average IGM transmission as a function of distance from galaxies $T(r)$ using 
\begin{equation}
\langle T(r) \rangle = \overline{\Sigma_{x,i} T (r(z_{Ly\alpha}(i),z_{x,\rm s}))}
\end{equation}
summing over each galaxy $x$ and spectral pixel $i$ pair in the Ly$\alpha$ forest and then averaging. $z_{Ly\alpha}(i)$ is the Ly$\alpha$ absorption redshift of the spectral pixel $i$, and $z_{x,\rm s}$ is the spectroscopic redshift of the galaxy $x$. $r(z_{Ly\alpha}(i),z_{x,\rm spec})=\sqrt{r_{\perp}^2+r_{\parallel}^2}$, where $r_{\perp}=\theta D_{A}(z_{x,\rm s})\times(1+z_{x,\rm s})$ is the transverse comoving distance between the galaxy $x$ and quasar sightline at redshift $z_{x,\rm s}$, $\theta$ is the angular separation between the galaxy $x$ and the quasar, and $D_{A}(z_{x,\rm s})$ is the angular diameter distance at $z_{x,\rm s}$. $r_{\parallel}=|D_c(z_{Ly\alpha}(i))-D_c(z_{x,\rm s})|$ is the comoving distance between the redshifts of spectral pixel $i$ and the galaxy $x$ along the line of sight. We use the bootstrap resampling to generate 1000 realizations of the galaxy sample, compute the IGM-galaxy cross-correlation function of each realization, and estimate the $1\sigma$ confidence intervals of the cross-correlation function. Because we only include one single field in the IGM-galaxy cross-correlation function, the uncertainty estimated from bootstrapping does not account for cosmic variance, as previously discussed in \citet{Kakiichi2025arXiv}.

\begin{figure}
    \centering
    \includegraphics[width=0.95\linewidth]{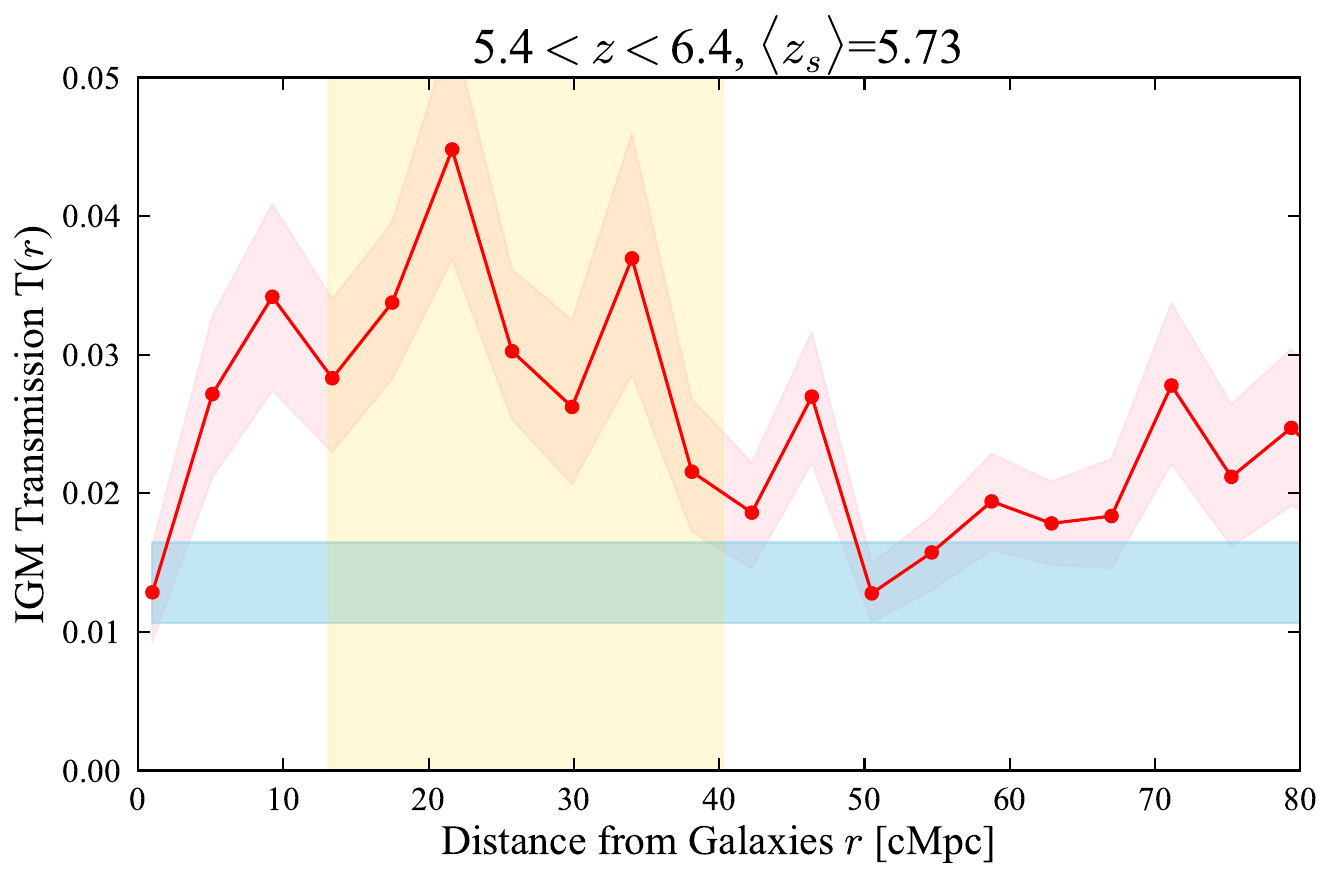}
    \caption{The IGM-galaxy cross-correlation function of spectroscopically confirmed sources at $5.4<z<6.4$. The average redshift of sources is $z=5.73$ and the average IGM transmission ($1\sigma$ range) at $z=5.73$ is denoted by the blue shaded area \citep{Yang2020ApJ}. The pink colored region denotes the $1\sigma$ confidence intervals estimated through bootstrapping. The 1$\sigma$ confidence intervals of the ionizing photon mean free path ($\lambda_{\rm mfp}=22.0_{-8.9}^{+18.2}~{\rm cMpc}$) at $z=5.65$ are denoted by the vertical yellow region \citep{Zhu2023ApJ}.}
    \label{fig:igm_galaxy_ccf}
\end{figure}

In Figure \ref{fig:igm_galaxy_ccf}, the cross-correlation function between IGM transmission and galaxies is plotted in red, and the $1\sigma$ confidence intervals from bootstrapping are shown in the pink region. Given the maximum transverse distance between galaxies and the quasar is only $\sim8$~cMpc (see Figure \ref{fig:los_plot}), the IGM transmission at scales $\gtrsim10~$cMpc from galaxies is mainly probed by the radial direction \cite[see also][]{Garaldi2025OJAa}. As such, the IGM-galaxy cross-correlation function at large scales is likely dominated by line-of-sight fluctuations and cosmic variance. The average redshift of sources used in the IGM-galaxy cross-correlation function is $5.73$. Compared with the average IGM transmission at $z=5.73$ \citep{Yang2020ApJ}, the IGM transmission is suppressed within $3-4$~cMpc from sources. At $r\sim10-40$~cMpc, there is an excess of the IGM transmission above the average IGM transmission at $z=5.73$. Notably, the scale of the excess IGM transmission at $10-40$~cMpc is comparable to the ionizing photon mean free path $\lambda_{\rm mfp}=22.0_{-8.9}^{+18.2}~{\rm cMpc}$ at $z=5.65$ \citep{Zhu2023ApJ}. At a larger distance ($\gtrsim50$cMpc), the IGM transmission around galaxies decreases and approaches the average IGM transmission at $z=5.73$. The IGM-galaxy cross-correlation function also displays fluctuations on scales of $\sim10-15$~cMpc. These small-scale fluctuations could be caused by the existence of individual strong spikes at $z\sim5.5$, which have separations around $15~$cMpc. 

\begin{figure*}
    \centering
    \includegraphics[width=0.9\linewidth]{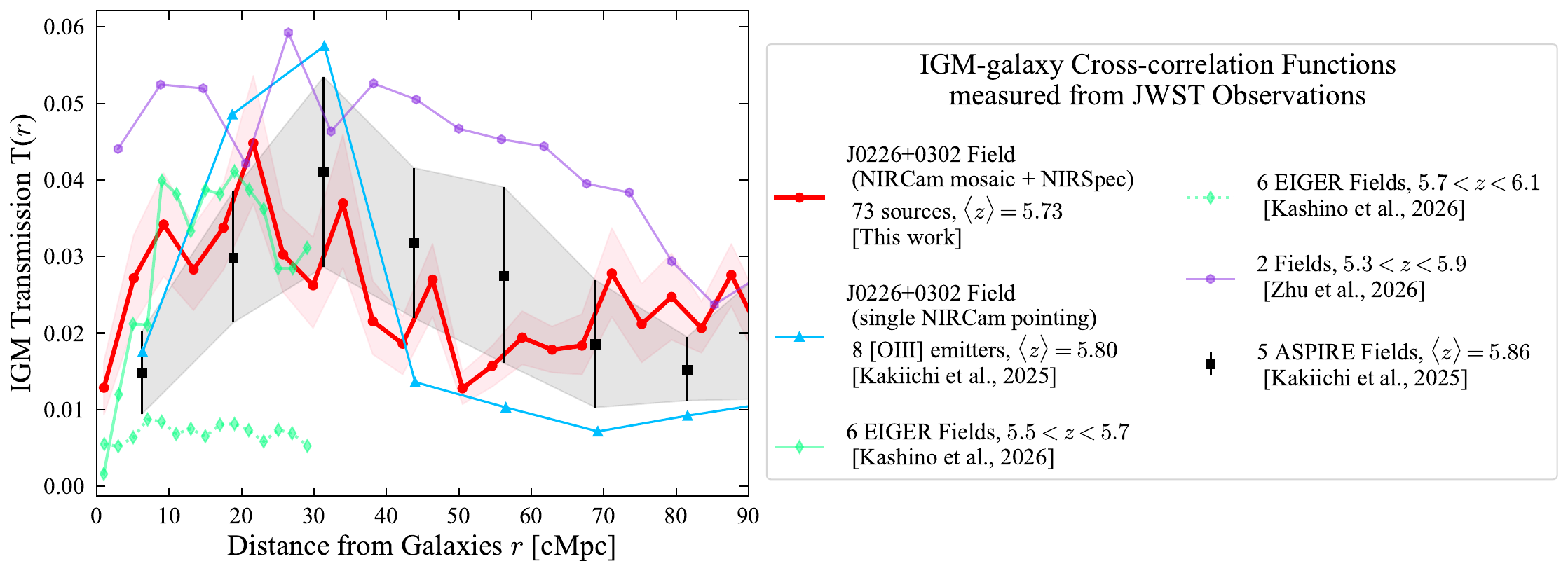}
    \caption{The IGM-galaxy cross-correlation function measured in the J0226 field with 73 sources selected from JWST cycle 1 and cycle 2 program (red). The IGM-galaxy cross-correlation function based on eight \oiii\ emitters selected from ASPIRE program is shown in blue solid line, and the IGM-galaxy cross-correlation function measured from 5 ASPIRE quasar fields is shown in a gray shaded region \citep{Kakiichi2025arXiv}. The IGM-galaxy cross-correlation functions from 6 EIGER quasar fields are plotted in a green solid line ($5.5<z<5.7$) and a green dotted line ($5.7<z<6.1$) \citep{Kashino2025arXiv}.  The IGM-galaxy cross-correlation function measured from two most transparent quasar sightlines at $z=5.7$ is plotted in purple (adapted from \citealt{Zhu2025arXiv}).}
    \label{fig:compare_with_kakiichi}
\end{figure*}

In Figure \ref{fig:compare_with_kakiichi}, we compare our result with the observed IGM-galaxy cross-correlation functions measured from other JWST observations, including ASPIRE \citep{Kakiichi2025arXiv}, EIGER \citep{Kashino2023ApJ,Kashino2025arXiv} and GO 4092 \citep{Becker2023jwst,Zhu2025arXiv}. 
The IGM-galaxy cross-correlation function reported in \citet{Kakiichi2025arXiv} were measured from five quasar fields, including the J0226 field, observed by single NIRCam pointings from ASPIRE program \citep{Wang2023ApJ}, covering two close $2.2'\times2.2'$ FoVs. 
\citet{Kashino2025arXiv} reported the IGM-galaxy cross-correlation function measured from six EIGER quasar fields up to $30$~cMpc. Each EIGER field was observed with a $2\times2$ NIRCam mosaic, covering a FoV of $3'\times6'$. \citet{Zhu2025arXiv} measured the IGM-galaxy cross-correlation function in two quasar fields with the most transparent IGM transmission at $z\sim5.7$ (GO 4092, PI: G. Becker), and each quasar field was observed by a $4\times2$ NIRCam mosaic, corresponding to a FoV of $10'\times7.5'$. 

By comparing the IGM-galaxy cross-correlation functions both measured in the J0226 field, but with different sizes of the FoV (single pointing (blue) -- \citet{Kakiichi2025arXiv}, mosaic (red) -- this work), we find two IGM-galaxy cross-correlation functions show good agreement. The IGM-galaxy cross-correlation function in \citet{Kakiichi2025arXiv} has a more enhanced IGM transmission, and the peak of the excess IGM transmission also appears at a larger scale than the IGM-galaxy cross-correlation function measured in this work. This could be attributed to the fact that only eight \oiii\ emitters were detected by \citet{Kakiichi2025arXiv}, and therefore the IGM-galaxy cross-correlation function might not be precisely sampled in a single quasar field with only eight objects. 

 We find the overall shape of the IGM-galaxy cross-correlation function measured in this work is also similar to the IGM-galaxy cross-correlation function of five ASPIRE fields presented by \citet{Kakiichi2025arXiv}, but the peak in the excess transmission is closer to the galaxy than the IGM-galaxy cross-correlation function of five ASPIRE fields \citep{Kakiichi2025arXiv}. Interestingly, we find the IGM-galaxy cross-correlation function is very consistent with the IGM-galaxy cross-correlation function within 30 cMpc at $5.5<z<5.7$ measured from 91 \oiii\ emitters in six EIGER quasar fields \citep{Kashino2025arXiv}. This implies that the discrepancy between EIGER and ASPIRE IGM-galaxy cross-correlation functions may arise from cosmic variance, but different survey depth and area of EIGER and ASPIRE may also play a role in causing the difference between IGM-galaxy cross-correlation functions. 

 The IGM-galaxy cross-correlation function presented by \citet{Zhu2025arXiv} has a higher transmission than other IGM-galaxy cross-correlation functions, and the scales of the excess transmission also extend to larger scales ($\sim80~{\rm cMpc}$) around galaxies. This is mainly due to the fact that two quasar sightlines in \citet{Zhu2025arXiv} exhibit the most transparent Ly$\alpha$ transmission at $z=5.7$. This large scatter of the IGM-galaxy cross-correlations also reflect the effect of cosmic variance.

\subsection{Comparison with THESAN Simulations}

\begin{figure*}
    \centering
    \includegraphics[width=0.95\linewidth]{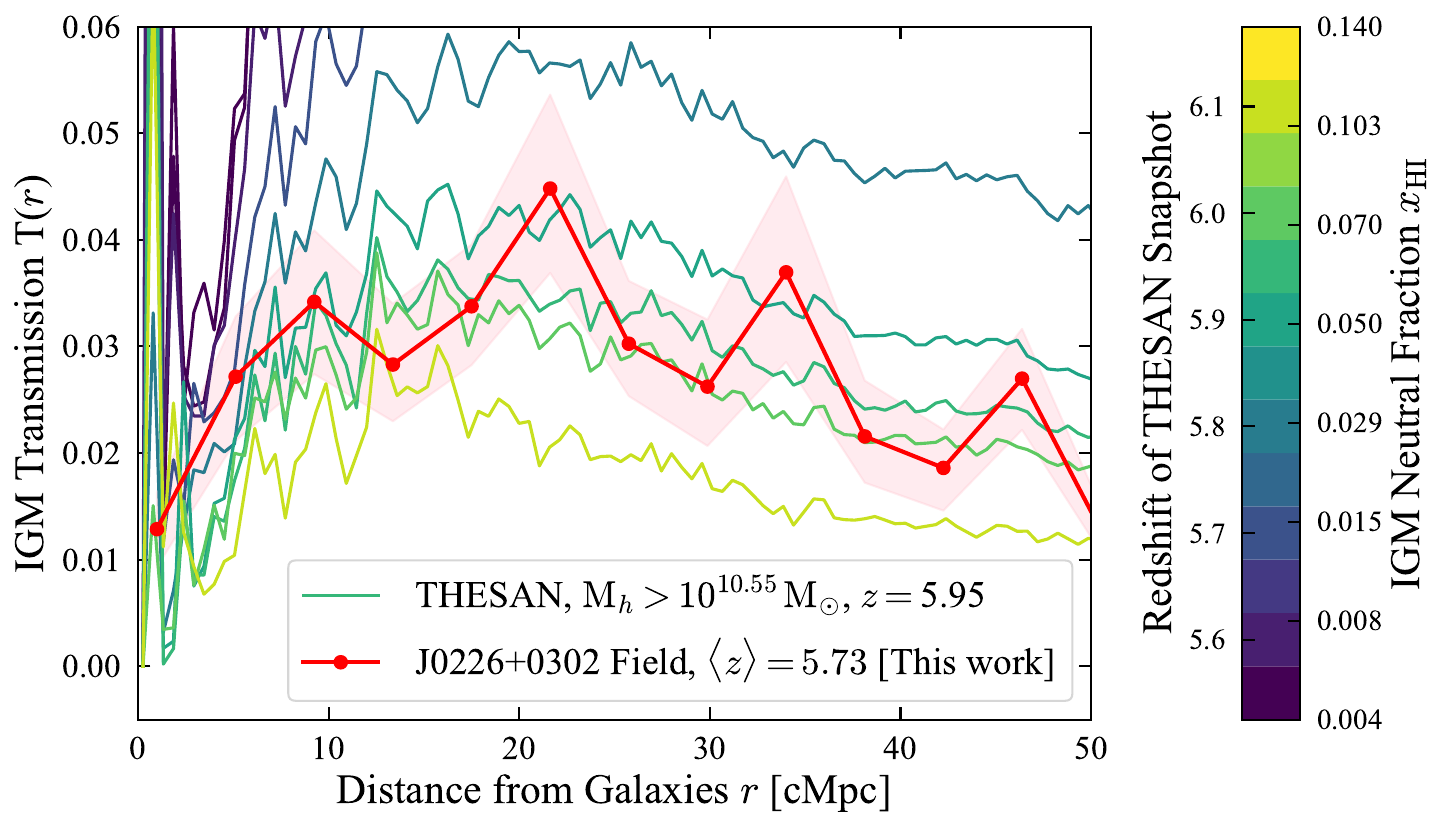}
    \caption{The IGM-galaxy cross-correlation function extracted from THESAN snapshots, using galaxies with halo masses ${\rm M}_ {h}>10^{10.55}\;\!{\rm M_{\odot}}$ (\citealt{Huang2026arXiv}, see also \citealt{Eilers2024ApJ}). The redshift of each THESAN snapshot is color-coded. The IGM galaxy cross-correlation function measured in the J0226 field is plotted in red.}
    \label{fig:thesan}
\end{figure*}

Figure \ref{fig:thesan} shows the IGM-galaxy cross-correlation function from THESAN simulations \citep{Kannan2022MNRAS,Garaldi2022MNRAS,Garaldi2024MNRAS,Smith2022MNRAS}. 
The THESAN simulations are a suite of radiation-magneto–hydrodynamical cosmological simulations designed to simultaneously model structure formation and cosmic reionization. They combine the IllustrisTNG galaxy formation model \citep{Weinberger2017MNRAS, Pillepich2018MNRAS} with a model for cosmic dust \citep[based on][]{McKinnon2017MNRAS}, multi-frequency radiative transfer and non-equilibrium primordial thermochemistry. These features make THESAN ideally suited for interpreting the high-redshift IGM-galaxy connection. Here we employ the flagship THESAN-1 box, covering a volume of $(95.5 \, \mathrm{cMpc})^3$ with sufficient resolution to capture atomic cooling haloes. In THESAN-1, all halos contribute to reionization and the ionizing photon escape fraction $f_{\rm esc}$ decreases as the halo masses increases, except for the most massive halo (${\rm M}_{h}>10^{11}~{\rm M_{\odot}}$), and the effective escape fraction averaging over all ionizing sources is around $6\%$ at $z=6$ \citep{Yeh2023MNRAS}.

We extract the IGM-galaxy cross-correlation function from THESAN-1 in redshift snapshots in the redshift range $z\sim5.5-6.2$. The redshifts snapshots are evenly spaced in cosmic time, and the redshift spacing is roughly ${\rm d}z=0.05$. We only select galaxies with host halo mass ${\rm M}_{h} > 10^{10.55}~{\rm M_{\odot}}$ (host halo mass of ASPIRE \oiii\ emitters, \citealt{Huang2026arXiv}, see also \citealt{Eilers2024ApJ}), when extracting the IGM-galaxy cross-correlation functions from THESAN-1. The size of THESAN simulation boxes is 95.5~cMpc, as such, we only compare the IGM-galaxy cross-correlation function within 50~cMpc. We find that the observed IGM-galaxy cross-correlation function in the J0226 field roughly matches the THESAN IGM-galaxy cross-correlation functions at $z\sim5.9$--$6.0$. These redshift snapshots have an average IGM neutral fraction $x_{\rm HI}\sim5\%-7\%$. This suggests the IGM-galaxy cross-correlation function measured from the J0226 field is consistent with the reionization scenario in which the IGM is not fully ionized at $z<6$, although a single field cannot constrain $x_{\rm HI}$ independently.
In addition, the IGM-galaxy cross-correlation function in the J0226 field displays a more rapid drop in transmission with distance than THESAN simulations at $\gtrsim30$~cMpc. 

Several factors could contribute to this discrepancy. First, the box size (95.5~cMpc) of current THESAN simulations is limited, which may not be large enough to capture the IGM transmission from galaxies at larger scales \citep[see e.g. Fig.~5 in][]{Conaboy2025}. 
Moreover, the overall shape of the IGM-galaxy cross-correlation function can also depend on the types of reionization sources \citep{Garaldi2022MNRAS}, and spectral energy distribution of reionization sources \citep{Basu2025arXiv}. Different reionization source models can also significantly change the fluctuations of the IGM temperature \citep{Asthana2025MNRAS}, which leads to different relations between IGM transmission and reionization sources \citep{Gangolli2025JCAP,Garaldi2025OJAb}. Further simulations are needed to quantify the impact of those factors on the IGM-galaxy cross-correlation functions.

\subsection{Metal Absorber Neighboring Galaxies}\label{subsec:metal}

In Figure \ref{fig:los_plot}, we plot the redshift of metal absorbers identified from the VLT/X-Shooter NIR spectrum, reported in \citet{Davies2023MNRAS}. 
We show the redshifts of \ion{C}{4}, \ion{Mg}{2}, and \ion{O}{1} absorbers that trace multiphase gas in the circumgalactic medium (CGM) \citep{Tumlinson2017ARAA}.

\citet{Finlator2020MNRAS} investigate the expected galaxy environments around \ion{C}{4} absorbers at $z>5$ through simulations, and find galaxies could correlate with \ion{C}{4} absorbers up to 300~pkpc. We find that among \ion{C}{4} absorbers at $z>5.4$, only the \ion{C}{4} absorber system at $z\sim5.9$ has two neighboring galaxies detected within 300~pkpc within our detection limit. The \ion{C}{4} absorber system at $z\sim5.9$ is composed of two \ion{C}{4} absorbers, one at $z=5.8987$ and the other at $z=5.9024$. The impact parameter between the \ion{C}{4} absorbers and the neighboring galaxies is 175~pkpc and 95~pkpc, respectively. The \ion{C}{4} absorber at $z=5.8987$ is a strong absorber with a column density ${\rm log N_H}=13.68$ \citep{Davies2023MNRAS}, while the other \ion{C}{4} absorbers without neighboring galaxies detected within 300~pkpc all have a column density ${\rm log N_H} < 13.50$. Our finding is in general agreement with \citet{Finlator2020MNRAS}, who finds the \ion{C}{4}-galaxy cross-correlation strengthens with both increasing galaxy luminosity and increasing absorber strength. 

Among \ion{Mg}{2} absorption systems, no neighboring galaxy is detected within 300~pkpc. An overdensity of six galaxies is detected within 1~pMpc from the $z=5.426$ \ion{Mg}{2} absorber. 
For the \ion{O}{1} and \ion{Mg}{2} absorption system at $z=6.06$, the closest galaxy is detected with an impact parameter of 388~pkpc. Through THESAN-ZOOM simulations \citep{Kannan2025OJAp}, \citet{Pruto2026MNRAS} find that low-mass galaxies $M_{\ast}<10^{8}\;\!{\rm M_{\odot}}$ are often found near \ion{O}{1} absorber. These metal absorption systems may serve as tracers of faint, low-mass galaxies, below our detection limits \cite[$M_{\ast}\sim10^{8.3}~{\rm M_{\odot}}$, ][]{Champagne2025ApJb}. A systematic search of neighboring galaxies near metal absorption system of different species will provide constraints on their host halo mass \cite[e.g.,][]{Wu2023ApJ}, and further paving the way for applying metal absorbers as galaxy tracers to study the IGM transmission around faint galaxies \cite[e.g.,][]{Becker2006ApJ,Meyer2019MNRAS}.

\section{Ionizing Photon Production Efficiency $\xi_{\rm ion}$}\label{sec:ion}

\begin{figure*}
    \centering
    \includegraphics[width=0.45\linewidth]{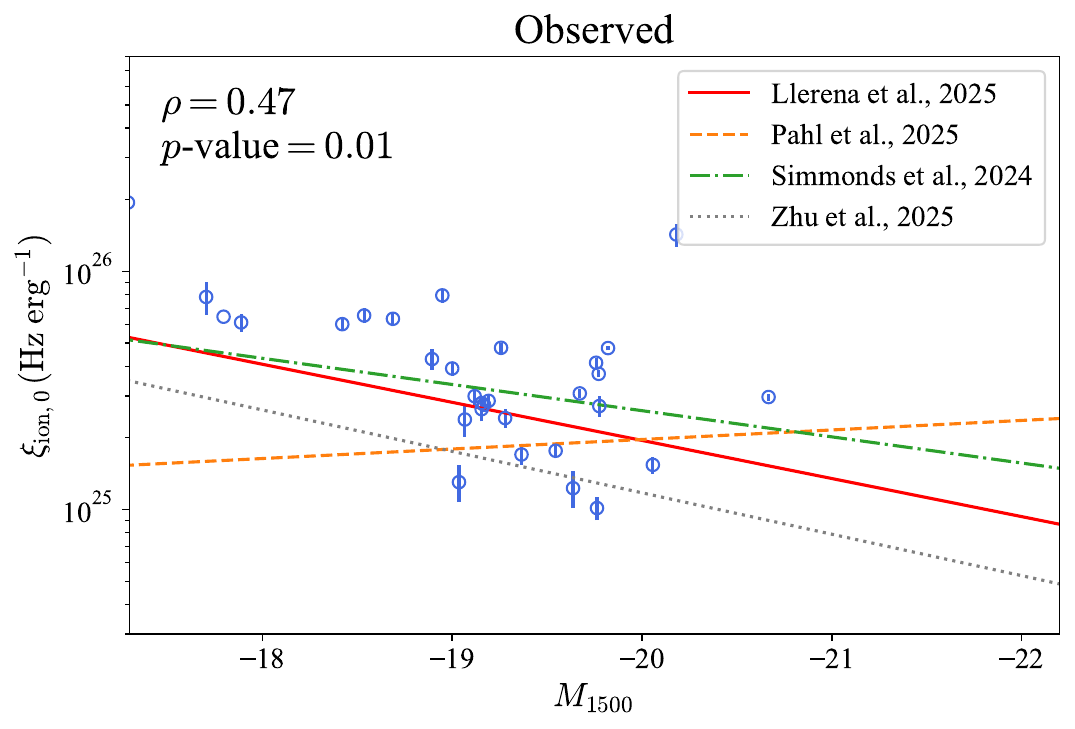}
    \includegraphics[width=0.45\linewidth]{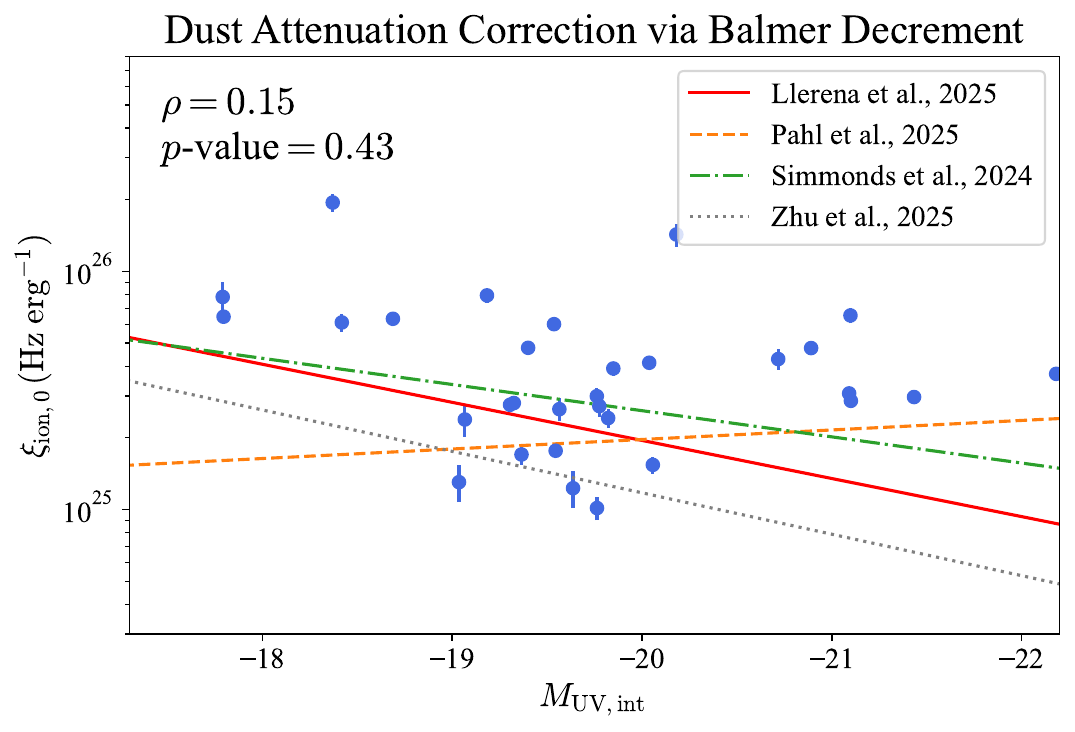}
    \caption{The correlation between ionizing photon production efficiency $\xi_{\rm ion}$ and $M_{\rm UV}$. The left panel shows the observed $\xi_{\rm ion}$ and $M_{\rm 1500}$, and the right panel shows the dust-corrected $\xi_{\rm ion}$ and $M_{\rm UV}$, using the Balmer decrement. In both panels, the best-fit $\xi_{\rm ion}-M_{\rm UV}$ relations from \citet{Simmonds2024MNRAS} (red solid), \citet{Llerena2025AA} (orange dashed), \citet{Pahl2025ApJ} (green dash-dotted), and \citet{Zhu2025ApJ} (gray dotted).}
    \label{fig:xiion}
\end{figure*}

For sources with H$\alpha$ detection at $\gtrsim3\sigma$ levels, we measure their ionizing photon production efficiency $\xi_{\rm ion}$ using $\xi_{\rm ion}=N(H^{0})/L_{\rm UV,int}$, where $N(H^{0})$ is the ionizing photon production rate and $L_{\rm UV,int}$ is the unattenuated UV luminosity at $1500~{\rm \AA}$. We remove broad line H$\alpha$ emitters during the $\xi_{\rm ion}$ calculations. 
We estimate the ionizing photon production rate using $N(H^{0})=7.35\times10^{11}\times L_{\rm H\alpha,int}/(1-f_{\rm esc})$, assuming a gas temperature of $10^{4}~$K and Case B \citep{Leitherer1995ApJS}, where $L_{\rm H\alpha,int}$ is the unattenuated H$\alpha$ luminosity and $f_{\rm esc}$ is the ionizing photon escape fraction. 
We calculate the ionizing photon production efficiency $\xi_{\rm ion,0}$ by assuming $f_{\rm esc}=0$, through $\xi_{\rm ion,0}=7.35\times10^{11}\times L_{\rm H\alpha,int}/L_{\rm UV,int}$. Any non-zero $f_{\rm esc}$ will yield a higher $\xi_{\rm ion}$ than $\xi_{\rm ion,0}$. 
 
To measure the unattenuated H$\alpha$ luminosity and UV luminosity, we adopt 31 sources with H$\beta$ detection at $\gtrsim3\sigma$ levels, and calculate the Balmer decrement to determine the level of the dust extinction. We assume an intrinsic H$\alpha$ to H$\beta$ flux ratio of 2.86 \citep{Dopita2003adu}, adopt the dust reddening curve in \citet{Calzetti2000ApJ}, and calculate the color excess $E(B-V)$ using $E(B-V)=1.97\times{\rm log_{10}}\;\![\frac{({\rm H}\alpha/{\rm H}\beta)_{\rm obs}}{2.86}]$ \citep{Dominguez2013ApJ}, where $({\rm H}\alpha/{\rm H}\beta)_{\rm obs}$ is the observed flux ratio of H$\alpha$ to H$\beta$. If $({\rm H}\alpha/{\rm H}\beta)_{\rm obs}$ is less than 2.86, we assume $E(B-V)=0$. We adopt $A_{\rm H\alpha}=3.33\times{E(B-V)}$ for H$\alpha$, and $A_{\rm UV}=10.33\times E_{s}(B-V)=4.55\times E(B-V)$ for the UV continuum \citep{Calzetti2000ApJ}, where $E_{s}(B-V)$ is the color excess of the stellar continuum, and calculate the unattenuated H$\alpha$ luminosity $L_{\rm H\alpha,int}$, unattenuated UV magnitude $M_{\rm UV, int}=M_{\rm 1500}-A_{\rm UV}$ and unattenuated UV luminosity $L_{\rm UV,int}$. 

We then calculate $\xi_{\rm ion,0}$ for the 31 sources with H$\beta$ detection. The left panel of Figure \ref{fig:xiion} shows the ``observed" $\xi_{\rm ion,0}$ as a function of $M_{\rm 1500}$, without dust extinction correction, and the right panel of Figure \ref{fig:xiion} shows $\xi_{\rm ion,0}$ as a function of the unattenuated $M_{\rm UV, int}$. In both panels of Figure \ref{fig:xiion}, we plot the $\xi_{\rm ion}-M_{\rm UV}$ relations reported by \citet{Simmonds2024MNRAS}, \citet{Llerena2025AA}, \citet{Pahl2025ApJ}, and \citet{Zhu2025ApJ}. The $\xi_{\rm ion}-M_{\rm UV}$ relations in \citet{Simmonds2024MNRAS} and \citet{Llerena2025AA} are measured from $z>4$ galaxies, while the relations in \citet{Pahl2025ApJ} and \citet{Zhu2025ApJ} include both cosmic noon and reionization-era galaxies, spanning $z\sim1-6$. We find that our observed and dust-corrected $\xi_{\rm ion}$ measurements are broadly consistent with the relations reported in the literature, although they may lie systematically above them, likely because our sample is predominantly emission-line selected. 

To examine the correlation between $\xi_{\rm ion}$ and $M_{\rm UV}$, we calculate the Spearman coefficient between ${\rm log_{10}}\xi_{\rm ion,0}$ and $M_{\rm UV}$. For the observed quantities, we find a Spearman coefficient $\rho=0.47$ ($p=0.01$) between the observed ${\rm log_{10}}\xi_{\rm ion,0}$ and $M_{\rm 1500}$, suggesting a significant correlation. After dust correction using the Balmer decrement, however, we find $\rho=0.15$ ($p=0.43$) between dust-corrected ${\rm log_{10}}\xi_{\rm ion,0}$ and $M_{\rm UV, int}$, indicating no significant correlation in our sample. To assess the robustness of this result given the limited sample size (31 sources), we bootstrap the sample 10,000 times and recompute the Spearman coefficient, obtaining $\rho=0.15_{-0.21}^{+0.19}$ ($68\%$ confidence interval). The $\rho$ distribution from bootstrapping spans both positive and negative values, and we therefore do not claim a significant correlation between ${\rm log_{10}}\xi_{\rm ion,0}$ and $M_{\rm UV, int}$, and a larger sample is required to fully probe the $\xi_{\rm ion,0}-M_{\rm UV}$ correlation.

\section{Transverse Proximity Effect of JWST AGN} \label{sec:discussion}

\subsection{IGM Optical Depth around AGN}

\begin{figure*}
     \includegraphics[width=0.3\linewidth]{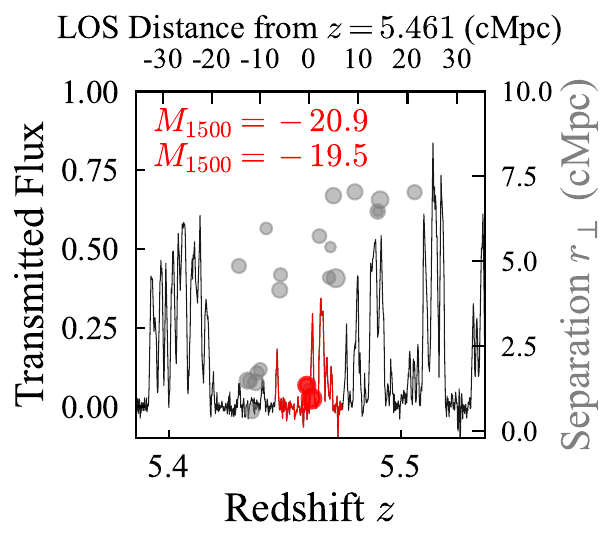}
     \includegraphics[width=0.3\linewidth]{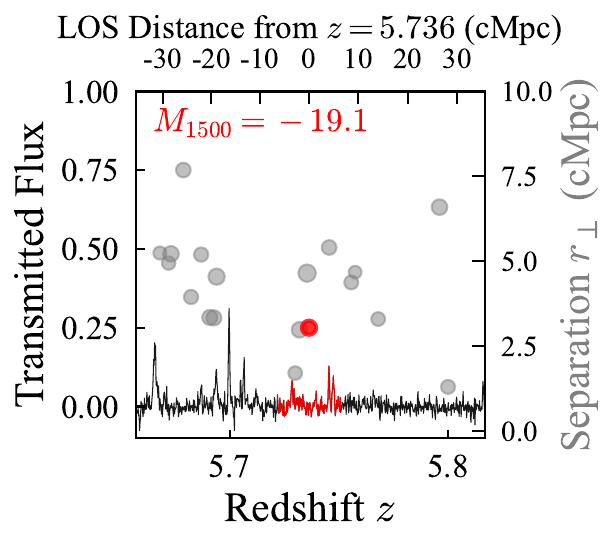}
    \includegraphics[width=0.3\linewidth]{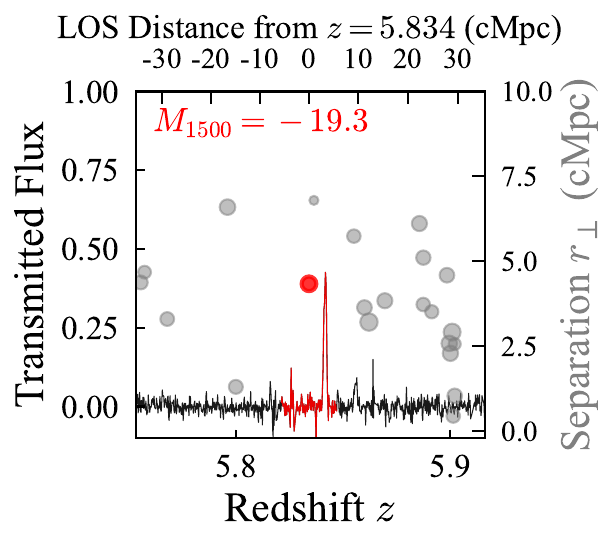}
    \caption{The Ly$\alpha$ forest within $25~h^{-1}~{\rm cMpc}$ of broad H$\alpha$ emitters. The broad H$\alpha$ emitters are denoted by red dots. The Ly$\alpha$ forest within $5~h^{-1}~{\rm cMpc}$ of the AGN is plotted in red. The UV magnitude $M_{\rm 1500}$ is annotated on each figure. Line-emitting galaxies with only narrow H$\alpha$ emission lines are denoted by grey dots. }
    \label{fig:zoom_in_forest_around_AGN}
\end{figure*}

As mentioned in Section \ref{sec:3}, we identify four AGN with broad H$\alpha$ emission lines from the NIRSpec/MSA observations. 
Figure \ref{fig:zoom_in_forest_around_AGN} shows the Ly$\alpha$ forest spectrum centered at the AGN redshifts. To study the transverse proximity effect of these AGN candidates, we first estimate the physical scale that we expect to detect the direct flux enhancement around JWST AGN based on their UV magnitude. We calculate the ``proximity zone" size of the AGN using 
\begin{equation}
R_{\rm eq}=\left(\frac{L_{912}\;\!\sigma_{0}}{4\pi h (\alpha_{\nu}^{\rm ion} + 2.75) \Gamma_{\rm bkg}(z)}\right)^{1/2}
\end{equation} \citep{Calverley2011MNRAS}, where $L_{912}$ is the AGN's monochromatic luminosity at 912~${\rm \AA}$, $\sigma_0$ is the photoionization cross-section for hydrogen at 912~${\rm \AA}$, $h$ is the Planck constant, $\alpha_{\nu}^{\rm ion}$ is the spectral slope of the Lyman Continuum spectrum ($f_{\nu}\propto \nu^{-\alpha_{\nu}^{\rm ion}}$), and $\Gamma_{\rm bkg}(z)$ is the photoionization rate of the ultravoilet background (UVB) at the AGN redshift $z$. $R_{\rm eq}$ defines the radius of a spherically symmetric ionized bubble powered only by the AGN, within which the photoionization rate from the AGN is greater than the UVB. We adopt the following assumptions when estimating $R_{\rm eq}$: 1) the AGN has 100\% LyC escape fraction, 2) the AGN provides 100\% of the observed UV light, 3) the AGN luminosity is persistent and the AGN lifetime is long enough to keep the ionizing front propagated and generate an ionized bubble (i.\,e.,\,AGN proximity zone) in the IGM \cite[][]{Bosman2020ApJ,Eilers2025arXiv}, and 4) no IGM attenuation within the AGN proximity zone. Given that the four AGN display a similar UV continuum slope $\beta$ to the quasar UV continuum slope (See Section \ref{sec:aper}), for each AGN, we adopt a broken power-law model for the AGN UV continuum and ionizing spectrum \citep{Shull2012ApJ}: a spectral index $\alpha_{\rm \lambda}$ of $-0.59$ at rest wavelength $\lambda<1000~{\rm \AA}$ (corresponding to $\alpha_{\nu}^{\rm ion}=1.41$ for the Lyman Continuum) and a spectral index $\alpha_{\rm \lambda}=-1.5$ at rest wavelength $\lambda>1000~{\rm \AA}$. We use this broken power-law model and the throughput of the F115W filter to fit the observed F115W magnitude of each AGN, and use the best-fit power-law model to calculate $L_{912}$. We adopt the UVB photoionization rate from \citet{Gaikwad2023MNRAS} based on the AGN redshift. The $R_{\rm eq}$ of four AGN is 2.2~cMpc at $z=5.460$, 4.2~cMpc at $z=5.462$, 2.2~cMpc at $z=5.736$, and 2.8 cMpc at $z=5.834$. For the two AGN at $z=5.46$, their separation is 1.3~cMpc, smaller than $R_{\rm eq}$ of either AGN. We therefore consider these two AGN at $z=5.46$ as an AGN pair, assume these two AGN are located within the same ionized bubble, and we set the midpoint of two AGN as the center the ionized bubble. We sum the $L_{912}$ of the AGN pair to calculate $R_{\rm eq}$ of the AGN pair, and we find $R_{\rm eq}=4.8~{\rm cMpc}$. Given the faint UV magnitudes, all four AGN have $R_{\rm eq}\lesssim5~{\rm cMpc}$. This motivates us to measure the IGM transmission within a small scale $\sim5~{\rm cMpc}$ of the AGN in order to search the transverse proximity effect of AGN. 

\begin{figure*}
    \centering
    \includegraphics[width=0.9\linewidth]
    {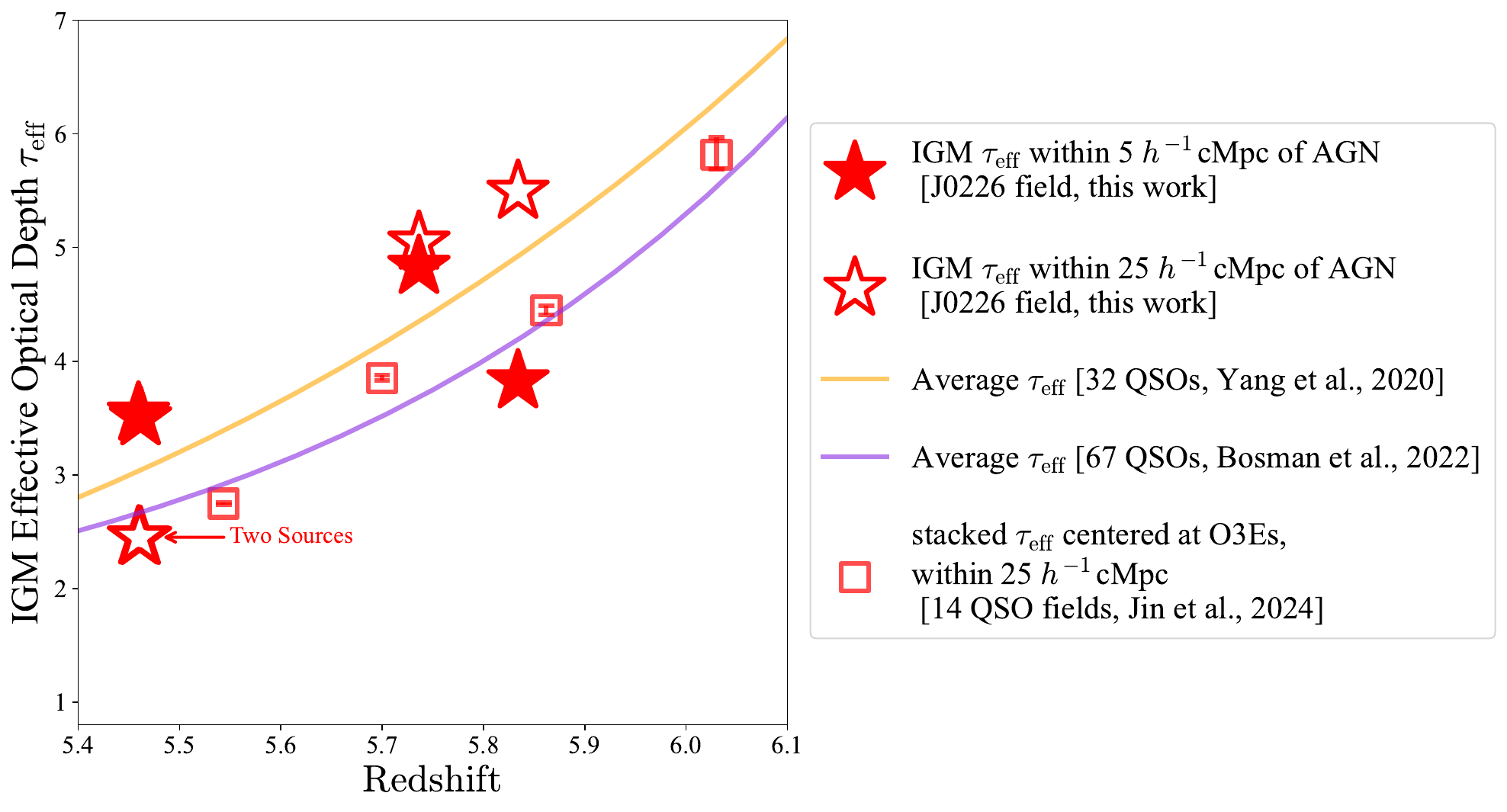}
    \caption{IGM effective optical depth ($\tau_{\rm eff}$) within  $5~h^{-1}\;\!{\rm cMpc}$ from the AGN is denoted by red filled stars. The $\tau_{\rm eff}$ within $25~h^{-1}\;\!{\rm cMpc}$ from the AGN is denoted by red open stars. The redshift evolutions of the average $\tau_{\rm eff}$ from \citet{Yang2020ApJ} and \citet{Bosman2022MNRAS} are shown by the orange and the purple lines. The $\tau_{\rm eff}$ of the stacked Ly$\alpha$ forest within $25~h^{-1}\;\!{\rm cMpc}$ of \oiii\ emitters is denoted by the red open squares \citep{Jin2024ApJ}.}
    \label{fig:igm_tau_around_an_AGN}
\end{figure*}

\begin{figure*}
    \centering
    \includegraphics[width=0.45\linewidth]{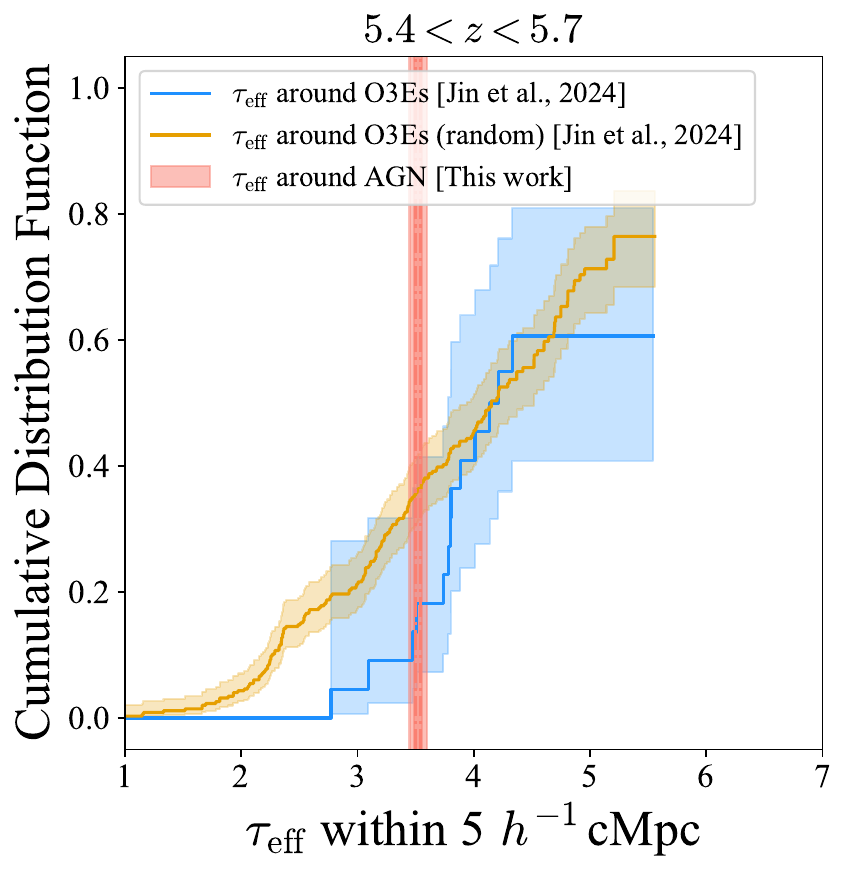}
    \includegraphics[width=0.45\linewidth]{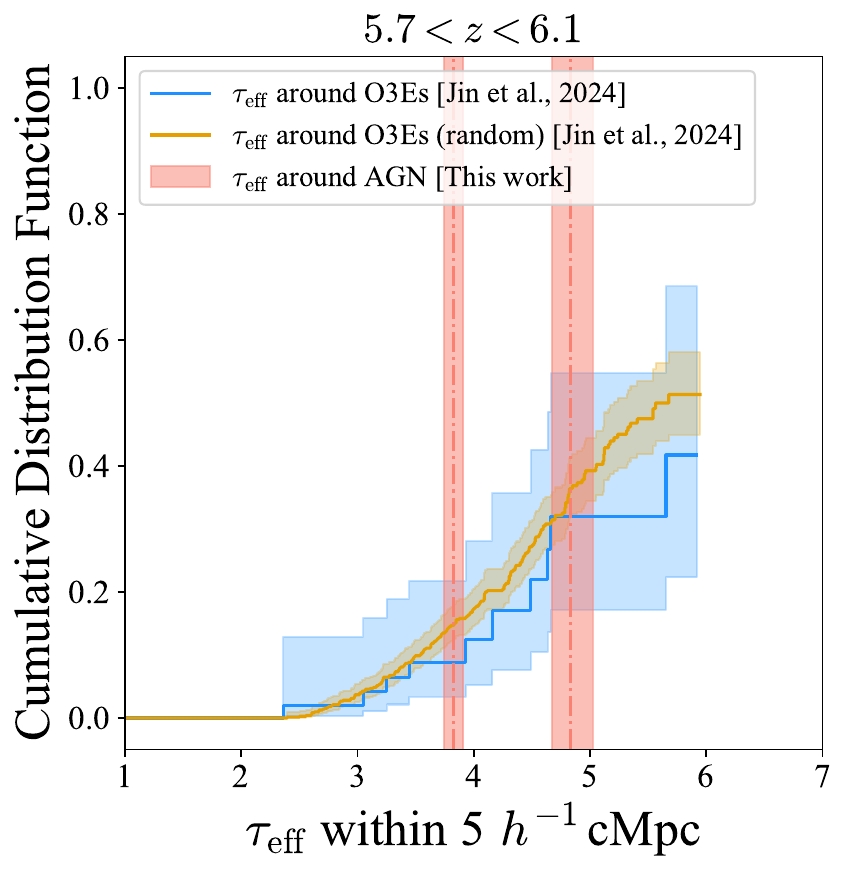}
    \caption{\textit{Left} - The blue shaded region denotes the cumulative distribution function (CDF) of IGM effective optical depth ($\tau_{\rm eff}$) within $5~h^{-1}\;\!{\rm cMpc}$ of \oiii\ emitters at $5.4<z<5.7$ estimated from Kaplan-Meier estimator \citep{Jin2024ApJ}. The yellow shaded region denotes the CDF of $\tau_{\rm eff}$ within $5~h^{-1}\;\!{\rm cMpc}$ of random \oiii\ emitters at $5.4<z<5.7$. The $\tau_{\rm eff}$ within $5~h^{-1}\;\!{\rm cMpc}$ of two AGN at $z=5.45-5.46$ are indicated by the vertical red regions. \textit{Right} - The blue shaded region denotes the CDF of $\tau_{\rm eff}$ within $5~h^{-1}\;\!{\rm cMpc}$ of \oiii\ emitters at $5.7<z<6.1$ calculated from the Kaplan-Meier estimator \citep{Jin2024ApJ}. The yellow shaded region denotes the CDF of $\tau_{\rm eff}$ within $5~h^{-1}\;\!{\rm cMpc}$ of random \oiii\ emitters at $5.7<z<6.1$ The $\tau_{\rm eff}$ within $5~h^{-1}\;\!{\rm cMpc}$ of two AGN at $z=5.736$ and $z=5.834$ are indicated by the vertical red regions. }
    \label{fig:cdf_tau}
\end{figure*}

Following \citet{Jin2024ApJ}, we measure the IGM transmission around AGN by placing a sphere centered at the AGN with a certain ``influence radius". We measure the IGM effective optical depth $\tau_{\rm eff}=-{\rm ln\langle T_{\rm Ly\alpha}\rangle}$, where $\langle T_{\rm Ly\alpha}\rangle$ is the average Ly$\alpha$ forest transmitted flux over the path length enclosed by the sphere centered at the AGN. We adopt an influence radius of $5\;\!h^{-1}~{\rm cMpc}$ to measure $\tau_{\rm eff}$ around the AGN. Figure \ref{fig:igm_tau_around_an_AGN} shows $\tau_{\rm eff}$ around the AGN within $5\;\!h^{-1}~{\rm cMpc}$. As a comparison, we also adopt an influence radius of $25\;\!h^{-1}~{\rm cMpc}$ to measure the large-scale IGM transmission around the AGN. We find for the two AGN at $z>5.7$, $\tau_{\rm eff}$ within $5\;\!h^{-1}~{\rm cMpc}$ of the AGN is lower than $\tau_{\rm eff}$ measured within $25\;\!h^{-1}~{\rm cMpc}$ of the AGN, indicating the IGM transmission is enhanced within $5\;\!h^{-1}~{\rm cMpc}$ of the IGM. But for the two AGN at $z<5.7$, $\tau_{\rm eff}$ within $5\;\!h^{-1}~{\rm cMpc}$ of the AGN is higher than large-scale IGM effective optical depth within $25\;\!h^{-1}~{\rm cMpc}$ of the AGN. 
However, such absorption within $5\;\!h^{-1}~{\rm cMpc}$ is similar to the Ly$\alpha$ absorption within $5\;\!h^{-1}~{\rm cMpc}$~cMpc previously found around \oiii\ emitters at $z<5.7$, potentially caused by the large-scale gas overdensity around \oiii\ emitters \citep{Kashino2023ApJ,Kashino2025arXiv,Jin2024ApJ,Kakiichi2025arXiv}. 

To better evaluate the level of IGM transmission near AGN, in Figure \ref{fig:cdf_tau}, we compare $\tau_{\rm eff}$ within $5\;\!h^{-1}~{\rm cMpc}$ of the AGN with the cumulative distribution function (CDF) of $\tau_{\rm eff}$ around \oiii\ emitters from \citet{Jin2024ApJ}. \citet{Jin2024ApJ} use the Kaplan-Meier estimator to estimate the CDF of $\tau_{\rm eff}$ around \oiii\ emitters within $5\;\!h^{-1}~{\rm cMpc}$ in 14 ASPIRE quasar fields, including both measurements and lower limits of $\tau_{\rm eff}$. The reconstructed CDF of $\tau_{\rm eff}$ does not reach unity because of a fraction of the $\tau_{\rm eff}$ measurements are lower limits (see blue and yellow shaded regions in Figure \ref{fig:cdf_tau}). \citet{Jin2024ApJ} compare the CDF of $\tau_{\rm eff}$ within $5\;\!h^{-1}~{\rm cMpc}$ of \oiii\ emitters (blue shaded regions in Figure \ref{fig:cdf_tau}) with a control CDF by randomly shuffling the locations of \oiii\ emitters in quasar fields (yellow shaded regions in Figure \ref{fig:cdf_tau}), and they report a significantly higher $\tau_{\rm eff}$ around \oiii\ emitters than the control sample at $5.4<z<5.7$. Given the limited number of AGN, a direct comparison between the CDF of $\tau_{\rm eff}$ around AGN and that around \oiii\ emitters is not feasible. Instead, we assess the location of the AGN $\tau_{\rm eff}$ measurements within the CDF of $\tau_{\rm eff}$ around \oiii\ emitters. We find that $\tau_{\rm eff}$ of two AGN at $z<5.7$ lie within the lowest 20\% of the CDF of $\tau_{\rm eff}$ within $5\;\!h^{-1}~{\rm cMpc}$ around \oiii\ emitters. For two AGN at $z>5.7$, we find their $\tau_{\rm eff}$ is located at the lowest 40\% of the CDF of $\tau_{\rm eff}$ around \oiii\ emitters. This indicates that the IGM transmission within $5\;\!h^{-1}~{\rm cMpc}$ is higher compared with median IGM transmission near \oiii\ emitters. This enhanced IGM transmission may trace the direct ionizing photon leakage from the AGN.

To assess the contribution from regions traced by the AGN to reionization on a larger scale, following \citet{Jin2024ApJ}, we adopt an influence radius of $25~h^{-1}\;\!{\rm cMpc}$ to measure the IGM optical depth around the AGN. Figure \ref{fig:igm_tau_around_an_AGN} shows the IGM effective optical depth around AGN, together with the average IGM effective optical depth evolution in \citet{Yang2020ApJ} and \citet{Bosman2022MNRAS}, and the stacked IGM effective optical depth measurements around \oiii\ emitters detected in 14 ASPIRE quasar fields, using an influence radius of $25~h^{-1}\;\!{\rm cMpc}$ \citep{Jin2024ApJ}. For two AGN at $z=5.46$, although the quasar Ly$\alpha$ forest at the source redshift shows gas absorption, when measuring the IGM transmission on a larger radius of $25~h^{-1}\;\!{\rm cMpc}$ from the AGN, the IGM transmission is higher than the average IGM transmission. In contrast, for two AGN at $z=5.736$ and $z=5.834$, the IGM is more opaque around them than the average IGM transmission. While for WFSS selected \oiii\ emitters in \citet{Jin2024ApJ}, the stacked Ly$\alpha$ transmission around \oiii\ emitters is significantly higher than the average IGM transmission when adopting an influence radius of $25~h^{-1}\;\!{\rm cMpc}$ to measure the effective optical depth centered at the \oiii\ emitters. This indicates that unseen ionizing sources clustered around the AGN might be different than the ionizing sources clustered around average \oiii\ emitters. Narrow H$\alpha$ emitters and broad-line AGN at $z\sim5$ have been found to have similar halo masses through auto-correlation function and cross-correlation functions \cite[e.\,g.,\,][]{Lin2025arXiva,Lin2025arXivb,Shuntov2025AA}, indicating that broad-line AGN reside in similar halos as star-forming galaxies. It is thus intriguing to witness that the large-scale IGM transmission around AGN might be different than the large-scale IGM transmission around star-forming galaxies traced by the majority of \oiii\ emitters. Nevertheless, our current sample of four AGN is very limited, therefore the cosmic variance could dominate the signal. A larger sample of broad-line AGN in the quasar fields is needed to further evaluate the IGM transmission around AGN in a statistical manner.

\subsection{IGM-AGN Cross-correlation Function}\label{sec:agn_igm_xcf}

\begin{figure}
    \centering
    \includegraphics[width=0.95\linewidth]{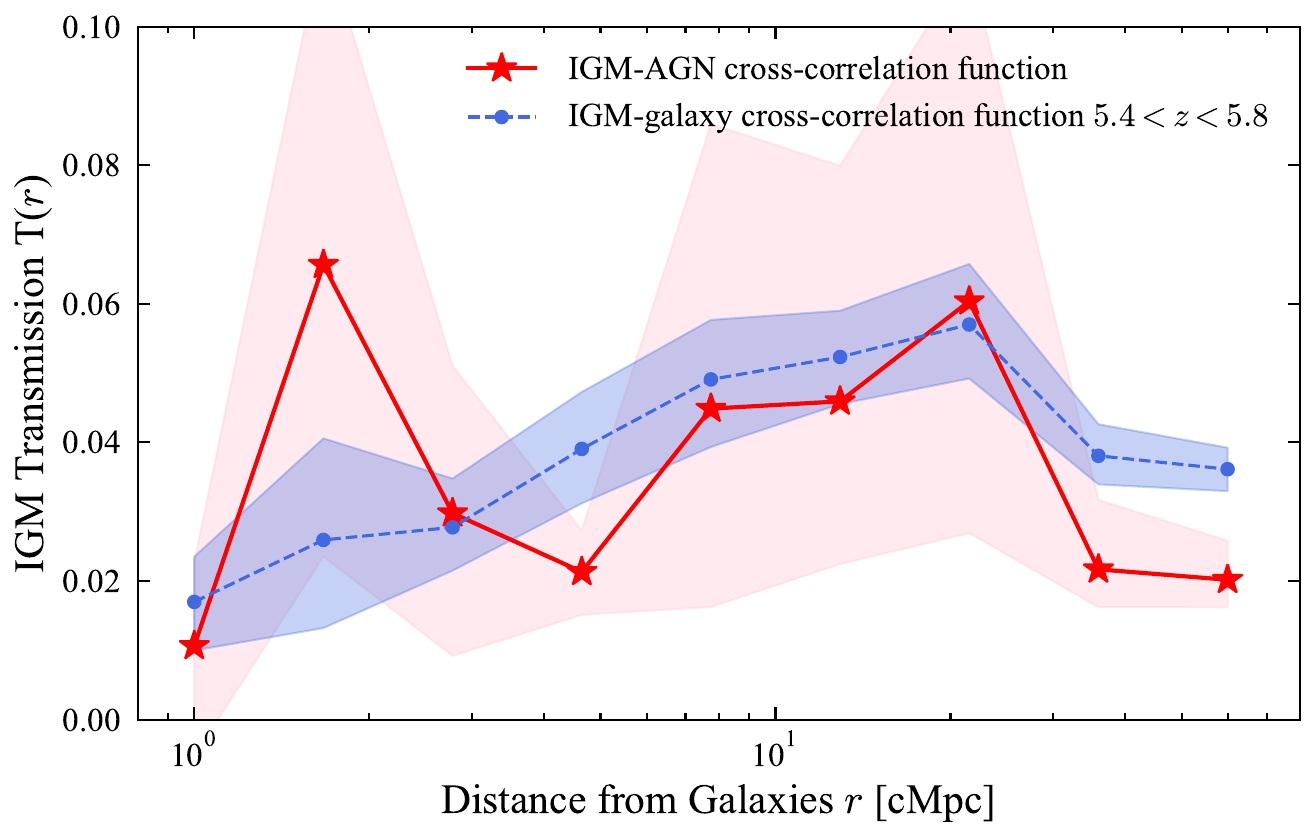}
    \caption{The IGM-AGN cross-correlation function around four broad-line AGN is denoted by the red line. The IGM-galaxy cross-correlation function at $5.4<z<5.8$ is denoted by the blue dashed line. The colored regions of each cross-correlation function denote the $1\sigma$ confidence intervals estimated from bootstrap resampling.}
    \label{fig:igm_agn_xcf}
\end{figure}

After examining the IGM effective optical depth of individual AGN, we measure the cross-correlation function between the quasar Ly$\alpha$ forest and four AGN to evaluate the average IGM transmission around the AGN. In Figure \ref{fig:igm_agn_xcf}, the IGM-AGN cross-correlation function is displayed in the red line. As a comparison, the IGM-galaxy cross-correlation function at $5.4<z<5.8$ is denoted by the blue dotted lines. The redshift bin of  $5.4<z<5.8$ covers the redshift range of four AGN. We calculate both IGM-AGN cross-correlation function and IGM-galaxy cross-correlation function with logarithmic bins of distance to focus on the IGM transmission at small-scales. We also estimate the $1\sigma$ uncertainty of the IGM-AGN cross-correlation function by bootstrap resampling of the AGN sample, as previously mentioned in Section \ref{sec:igm_galaxy_ccf}. 

Due to the limited number of AGN, the scatter in the IGM-AGN cross-correlation function is significant. However, a tentative excess transmission near the AGN is shown at 3~cMpc than the IGM-galaxy cross-correlation function at $5.4<z<5.8$. Such a feature has not been identified in IGM-galaxy cross-correlation functions at $z<5.8$ measured from previous JWST observations (see Figure \ref{fig:compare_with_kakiichi}). 
This excess transmission at small-scales in the IGM-AGN cross-correlation function is similar to IGM-AGN cross-correlation function at $z\sim2-3$, where the cross-correlation peaks at $5-6~h^{-1}\;\!{\rm cMpc}$, likely caused by the photoionization of AGN \citep{Momose2021ApJ}.

Because no current simulations predict the IGM-AGN cross-correlation function during reionization, we use an illustrative model to describe the tentative excess transmission and quantify the ionizing photon leakage from AGN. We model the total photoionization rate $\Gamma_{\rm HI}(r)$ at a distance $r$  from the AGN as $\Gamma_{\rm HI}(r) = \Gamma_{\rm HI}^{\rm AGN}(r) + \Gamma_{\rm HI}^{\rm GAL}(r)$ where $\Gamma_{\rm HI}^{\rm AGN}(r)$ is the photoionization rate of the AGN and $\Gamma_{\rm HI}^{\rm GAL}(r)$ is the photoionization rate of galaxies clustered around the AGN. We follow the method described in \citet{Bosman2020ApJ}. Because the speed of light is finite, the radiation of the AGN at the same three-dimensional distance $r$ can reflect the AGN activity at different lookback times (so-called the ``light-echo" effect, \citealt{Adelberger2004ApJ}), depending on both the line-of-sight distance $r_{\rm \parallel}$, and the transverse distance $r_{\perp}$ from the AGN, where $r=\sqrt{r_{\parallel}^2+r_{\perp}^2}$. The time-delay surface is given by $\Delta t(r_\parallel,r_\perp) = \frac{(r_\parallel^2 + r_\perp^2)^{1/2} + r_\parallel}{c}$, where $c$ is the speed of light. 
A negative $r_\parallel$ indicates a line of sight distance in front of the AGN. We can then model the photoionization rate of the AGN at $r_{\perp}$ and $r_{\parallel}$ with:
\begin{equation}
    \Gamma_{\rm HI}^{\rm AGN}(r)=\left(\int_{\nu_{912}}^{\infty}{\frac{f_{\rm esc} L_{\nu}[\Delta t(r_\parallel,r_\perp)]}{h\nu}\frac{\sigma(\nu)}{4\pi r^2}\;\! {\rm d}\nu}\right)\times e^{-r/\lambda_{\rm mfp}}, 
\end{equation}
where $f_{\rm esc}$ is the ionizing photon escape fraction of the AGN, 
$L_{\nu}[\Delta t(r_\parallel,r_\perp)]$ is the AGN ionizing luminosity at the ionizing photon's frequency $\nu$, given the time-delay surface $\Delta t(r_\parallel,r_\perp)$. $\sigma(\nu)=\sigma_{0}(\frac{\nu}{\nu_{912}})^{-2.75}$ is the photoionization cross-section as a function of ionizing photon frequency $\nu$, $\sigma_0$ is the photoionization cross-section for hydrogen at 912~${\rm \AA}$, and $\nu_{912}$ is the ionizing photon frequency at 912~${\rm \AA}$. $r$ is the distance from the AGN, and $\lambda_{\rm mfp}$ is the ionizing photon mean free path. 

\begin{figure*}[!htb]
    \centering
    \includegraphics[width=0.45\linewidth]{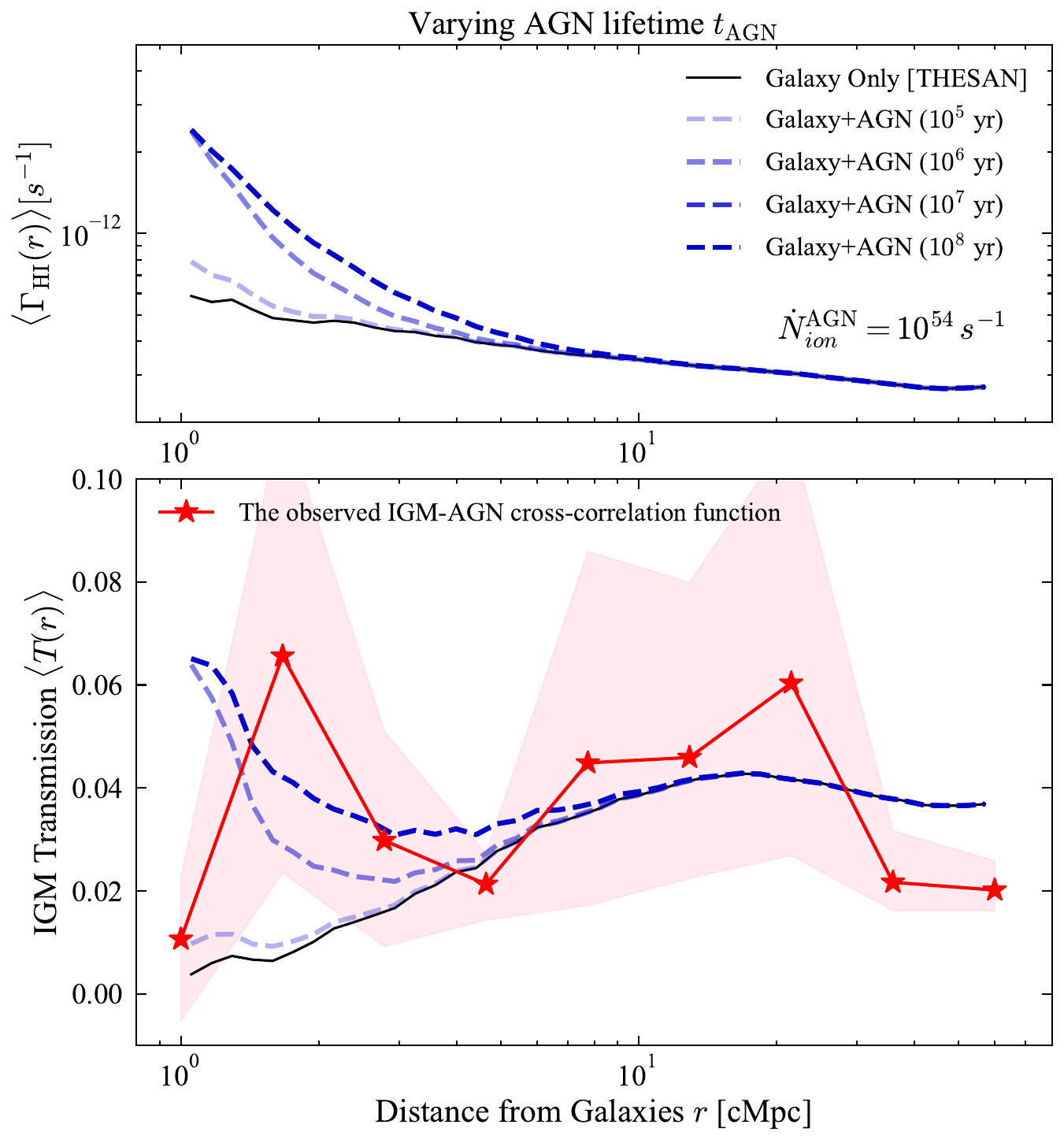}
    \includegraphics[width=0.45\linewidth]{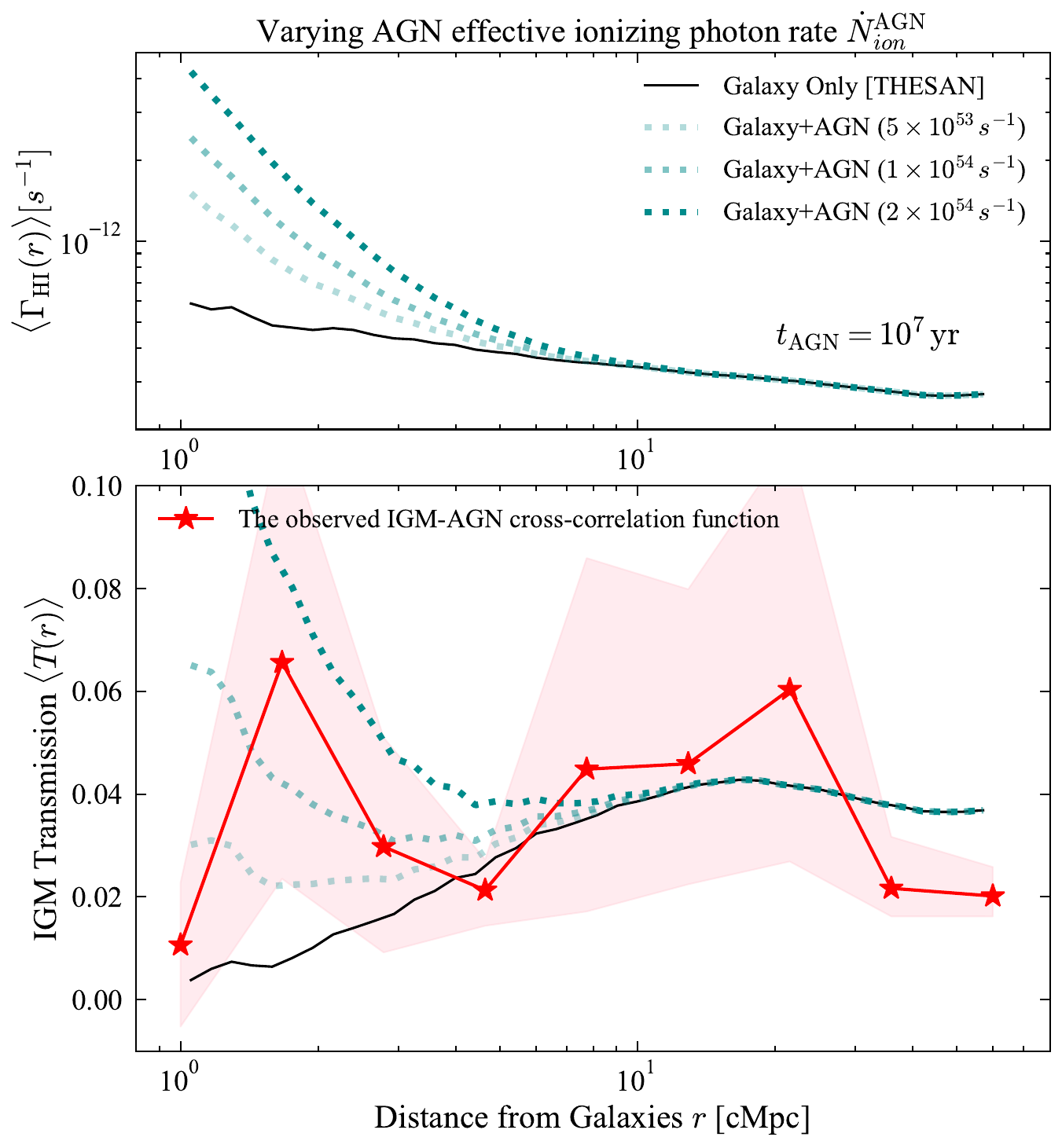}
    \caption{\textit{Top panels} -- The total photoionization rate $\Gamma_{\rm HI}$ (galaxy+AGN) as a function of distance $r$ from galaxies. The original photoionization rate in THESAN is plotted in black, and colored lines indicate the spherically-averaged photoionization rate by including AGN as the central ionizing sources with different AGN lifetimes $t_{\rm AGN}$(blue dashed lines in the top left panel) from $10^{5}$ to $10^{8}$ years, and with different AGN effective ionizing production rates $\dot{N}_{ion}^{\rm AGN}$ (cyan dotted lines in the top right panel) from $5\times10^{53}\;\!s^{-1}$ to $2\times10^{54}\;\!s^{-1}$. \textit{Bottom panels} -- The spherically-averaged IGM-AGN cross-correlation function, by scaling the photoionization rate. The observed IGM-AGN cross-correlation function is denoted by the red line. See Section \ref{sec:agn_igm_xcf} for more details. }
    \label{fig:agn_igm_model}
\end{figure*}

We use $\dot{N}_{ion}^{\rm AGN}[\Delta t(r_\parallel,r_\perp)]=\int_{\nu_{912}}^{\infty}{\frac{f_{\rm esc} L_{\nu}[\Delta t(r_\parallel,r_\perp)]}{h\nu}(\frac{\nu}{\nu_{912}})^{-2.75}\;\! {\rm d}\nu}$ to indicate the ``effective" ionizing photon rate that escapes from the AGN. As such, $\Gamma_{\rm HI}^{\rm AGN}(r_\parallel,r_\perp) = \sigma_{\rm 0}\frac{\dot{N}_{ion}^{\rm AGN}[\Delta t(r_\parallel,r_\perp)]}{4\pi r^2} e^{-r/\lambda_{\rm mfp}}$.
With a power-law ionizing spectrum ($L_{\nu}\propto\nu^{-\alpha_{\nu}^{\rm ion}}$), 
\begin{align}
    & \dot{N}_{ion}^{\rm AGN}[\Delta t(r_\parallel,r_\perp)]=\frac{f_{\rm esc} L_{\rm 912}[\Delta t(r_\parallel,r_\perp)]}{h(\alpha_{\nu}^{\rm ion} + 2.75)} \nonumber \\
    & =1.1\times10^{54}~{s^{-1}}(\frac{f_{\rm esc}}{100\%})\;\!(\frac{1.41+2.75}{\alpha_{\nu}^{\rm ion} + 2.75})\;\!(\frac{L_{\rm 912}[\Delta t(r_\parallel,r_\perp)]}{3\times10^{28}~{\rm erg\;\!s^{-1}\;\!Hz^{-1}}}).
\end{align} 

$\Gamma_{\rm HI}^{\rm AGN}(r_\parallel,r_\perp)$ is mainly determined by the AGN lifetime $t_{\rm AGN}$ and the AGN effective ionizing photon rate $\dot{N}_{ion}^{\rm AGN}$. We adopt $\lambda_{\rm mfp}=0.8~{\rm pMpc}$ \citep{Zhu2023ApJ} and calculate $\Gamma_{\rm HI}^{\rm AGN}(r_\parallel,r_\perp)$ for different $t_{\rm AGN}$ and $\dot{N}_{ion}^{\rm AGN}$. 
On the top panels of Figure \ref{fig:agn_igm_model}, we plot the spherically-averaged total photoionization rate $
\langle \Gamma_{\rm HI}(r) \rangle = \langle  \Gamma_{\rm HI}^{\rm AGN}(r) + \Gamma_{\rm HI}^{\rm GAL}(r)\rangle $ as a function of distance $r$ from the AGN. In both of top panels, the photoionization rate extracted from THESAN simulations $\Gamma_{\rm HI}^{\rm THESAN}$ at $z\sim5.83$ is denoted by the black line. In THESAN simulations, because the photoionization rate $\Gamma_{\rm HI}^{\rm THESAN}$ is dominated by stars, and AGN contribute less than 1\% of all ionizing photons at $z=6$ \citep{Yeh2023MNRAS}, we then assume $\Gamma_{\rm HI}^{\rm GAL}=\Gamma_{\rm HI}^{\rm THESAN}$. In the top left panel, we plot the total photoionization rate ($\Gamma_{\rm HI}^{\rm AGN}+\Gamma_{\rm HI}^{\rm THESAN}$) by assuming $\dot{N}_{ion}^{\rm AGN}=10^{54}\;\!s^{-1}$ and change the AGN lifetime $t_{\rm AGN}$ from $10^{5}~{\rm yr}$ to $10^{8}~{\rm yr}$. It is clear that the photoionization rate at $r>1~{\rm cMpc}$ of the AGN increases rapidly when increasing the AGN lifetime from $10^{5}~{\rm yr}$ to $10^{7}~{\rm yr}$. But when $t_{\rm AGN}$ is long enough ($\gtrsim10^{7-8}\;\!{\rm yr}$), the total photoionization rate does not change as $t_{\rm AGN}$ increases. On the top right panels of Figure \ref{fig:agn_igm_model}, we plot the photoionization rate by increasing the AGN ionizing photon production rate $\dot{N}_{ion}^{\rm AGN}$ from $5\times10^{53}~{s^{-1}}$ to $2\times10^{54}~{s^{-1}}$, with a given AGN lifetime $t_{\rm AGN}=10^{7}~{\rm yr}$.

We scale the IGM transmission in THESAN with photoionization rate in THESAN and AGN's photoionization rate. 
We adopt the fluctuating Gunn-Peterson approximation with $\tau_\alpha \propto \Delta_b^2\;\!\Gamma^{-1}_{\rm HI}\;\!\mathcal{T}_{\rm IGM}^{-0.7}$ \cite[e.\,g.,\,][]{Croft1998ApJ}, where $\tau_\alpha$ is the optical depth in the Ly$\alpha$ forest, $\Delta_{\rm b}$ is the overdensity, $\Gamma_{\rm HI}$ is the photoionization rate, and $\mathcal{T}_{\rm IGM}$ is the IGM temperature. We assume the same $\Delta_{\rm b}$ and $\mathcal{T}_{\rm IGM}$ around star-forming galaxies and AGN. With these assumptions, the Ly$\alpha$ forest optical depth $\tau_\alpha$ can be scaled with $\Gamma^{-1}_{\rm HI}$. We then scale the Ly$\alpha$ optical depth in THESAN with {$\tau_\alpha=\tau_{\rm \alpha,THESAN}\times\Gamma_{\rm HI}^{\rm THESAN}/(\Gamma_{\rm HI}^{\rm THESAN}+\Gamma_{\rm HI}^{\rm AGN})=\tau_{\rm \alpha,THESAN}\times1/(1+\Gamma_{\rm HI}^{\rm AGN}/\Gamma_{\rm HI}^{\rm THESAN})$, where $\tau_{\rm \alpha,THESAN}$ is the Ly$\alpha$ optical depth in THESAN, and $\Gamma_{\rm HI}^{\rm AGN}$ is the AGN photoionization rate. We then calculate the IGM transmission $T$ through $T=e^{-\tau_{\alpha}}$ to derive the model IGM-AGN cross-correlation function. We calculate the spherical mean of the IGM-AGN cross-correlation at each $r$. In the bottom panels of Figure \ref{fig:agn_igm_model}, we show the IGM-AGN cross-correlation function scaled with the photoionization rate. Increasing both $\dot{N}_{ion}^{\rm AGN}$ and $t_{\rm AGN}$ can create the excess transmission in the IGM-AGN cross-correlation function appearing at small scales $\lesssim5~$cMpc of the AGN. We find an AGN lifetime of $10^{6-8}\;\!{\rm yr}$ and an AGN effective photoionization rate of $\sim 5\times10^{53}-10^{54}\;\!s^{-1}$ can broadly reproduce the observed IGM-AGN cross-correlation function. Assuming a broken power-law for the AGN ionizing spectrum \citep{Shull2012ApJ}, the average $L_{912}$ is $2.8\times10^{28}~{\rm erg\;\!s^{-1}\;\!Hz^{-1}}$ for four AGN. To achieve an $\dot{N}_{ion}^{\rm AGN}\sim5\times10^{53}-10^{54}\;\!s^{-1}$, an average ionizing photon escape fraction $f_{\rm esc}\sim50\%-100\%$ is required for these four AGN. 

 \citet{Tang2025ApJ,Tang2026arXiv} and \citet{Ji2026arXiv} propose that ionizing photons may escape through the clumpy gas envelope near LRDs. 
 The IGM-AGN cross-correlation function provides a pathway to directly constrain the ionizing photon escape of LRDs, and further quantify their contribution to reionization. However, the current IGM-AGN cross-correlation function is measured in only a single quasar field and is thus strongly subject to cosmic variance. This limits both the amplitude and the characteristic scale of the excess transmission in the IGM-AGN cross-correlation function. Given that the ionizing spectra of such broad-line AGN also remain largely unconstrained (e.g., \citealt{Tang2025ApJ}, but see also \citealt{,WangBJ2026ApJ}), we emphasize that the $f_{\rm esc}\sim50\%$--$100\%$ range inferred from the IGM-AGN cross-correlation function should be regarded as indicative rather than a robust measurement. The characteristic scale of the excess transmission is also uncertain. Because we detect only four AGN over a limited range of UV magnitudes in a single quasar field, and the extent of the enhanced transmission region depends on the AGN luminosity and lifetime, the scale may vary between AGN and could extend beyond $\gtrsim10~$cMpc \cite[e.g.,][]{Gallerani2008MNRAS,Kakiichi2018MNRAS}. Larger samples of AGN in quasar fields, together with deep NIRSpec/MSA observations, are essential to measure the IGM-AGN cross-correlation function, detect the local IGM transmission enhancement near the AGN, and further determine the contribution of AGN to reionization as a population.

\section{Summary}\label{sec:summary}
In this paper, we present new JWST NIRCam and NIRSpec/MSA observations in a quasar field J0226$+$0302 at $z=6.5412$ to investigate the connections between IGM, galaxies, and AGN at the ending stages of reionization. We identify 73 line-emitting galaxies within the redshift range of the quasar Ly$\alpha$ forest at $5.4<z<6.4$ through emission-line detection in NIRCam/WFSS and NIRSpec/MSA observations. Our main findings are as follows: 

\begin{itemize}
    \item We measure the IGM-galaxy cross-correlation function, and find an excess IGM transmission at $10$--$40$~cMpc from galaxies. The scale of the excess IGM transmission is comparable to the ionizing photon mean free path at the average redshift of identified galaxies $z=5.7$. 
    \item The observed IGM-galaxy cross-correlation function is consistent with the THESAN IGM-galaxy cross-correlation function at $z\sim6$ with an average IGM neutral fraction of $5\%-7\%$ and an average ionizing photon escape fraction of $6\%$. 
    \item We measure the ionizing photon production efficiency $\xi_{\rm ion}$ of 31 sources with H$\beta$ detection. After performing dust correction using the Balmer decrement, we find no significant correlation between dust-corrected $\xi_{\rm ion}$ and $M_{\rm UV}$ in our sample. 
    \item Among 49 line-emitting galaxies observed by NIRSpec, we identify four sources with broad ($>1000~{\rm km\;\!s^{-1}}$) H$\alpha$ emission lines, corresponding to an AGN fraction of $(8\pm4)\%$. We detect a tentative excess IGM transmission within $5\;\!h^{-1}\;\!{\rm cMpc}$ of the AGN, seen in both the IGM effective optical depth around the AGN and the IGM-AGN cross-correlation function. We find that an AGN lifetime $\gtrsim10^{7}$ years and an ionizing photon escape fraction of $50\%-100\%$ can broadly reproduce the observed IGM-AGN cross-correlation function. 
\end{itemize}

The current IGM-AGN cross-correlation function still exhibits large uncertainty, due to cosmic variance and a limited number of AGN used in the measurements. Future JWST NIRCam and NIRSpec observations in quasar fields will significantly reduce the uncertainty in the IGM-AGN cross-correlation function and will fully characterize the scale of excess IGM transmission near the AGN,
further revealing the relation between IGM transmission and ionizing sources, and constraining the contribution of galaxies and AGN to reionization. 


\begin{acknowledgements}
XJ and FW acknowledge support from NSF award AST-2513040. K.K. acknowledges support from VILLUM FONDEN (71574) and the DAWN fellowship from the Cosmic Dawn Center, which is funded by the Danish National Research Foundation under grant no. 140.
    EG is supported by the JSPS KAKENHI grant ILR 23K20035. 
    M.V. gratefully acknowledges financial support from the Independent Research Fund Denmark via grant number 3103-00146 and from the Carlsberg Foundation (grant CF23-0417). HC thanks the support of the Natural Sciences and Engineering Research Council of Canada (NSERC), funding reference number RGPIN-2025-04798 and DGECR-2025-00136, and by the Department of Science at the University of Alberta Augustana Campus. SZ acknowledges support from the Chinese Academy of Sciences (no. E5295401).
    
    This work is based on observations made with the NASA/ESA/CSA James Webb Space Telescope. The data were obtained from the Mikulski Archive for Space Telescopes at the Space Telescope Science Institute, which is operated by the Association of Universities for Research in Astronomy, Inc., under NASA contract NAS 5-03127 for JWST. These observations are associated with programs \#2078 (ASPIRE) and \#3325. All the JWST data used in this paper can be found in MAST: \dataset[10.17909/zc6s-gb11]{http://dx.doi.org/10.17909/zc6s-gb11}. 

    This paper has benefited from grammar checks and writing improvements provided by Claude (Anthropic) and ChatGPT (OpenAI).
\end{acknowledgements}

\facilities{JWST, VLT (X-Shooter)}
\software{astropy \citep{2013A&A...558A..33A,2018AJ....156..123A,astropy:2022}, Numpy \citep{Numpy2020}, Matplotlib \citep{Matplotlib2007}, Scipy \citep{2020SciPy-NMeth}, JWST Calibration Pipeline \citep{2022zndo...7038885B}, msaexp \citep{Brammer2022zndo...7299500B}          }

\bibliography{sample7}{}
\bibliographystyle{aasjournalv7}

\suppressAffiliationsfalse
\allauthors


\end{document}